\documentclass[aps,prb,twocolumn,superscriptaddress,floatfix,longbibliography]{revtex4-2}
\usepackage[utf8]{inputenc}
\usepackage{amssymb}
\usepackage{amsmath}
\usepackage{mathtools}
\usepackage{bm}
\usepackage{graphicx}
\usepackage{multirow}
\usepackage{color}
\usepackage{hyperref}
\hypersetup{
    colorlinks=true,
    linkcolor=blue,
    citecolor=blue,
    urlcolor=blue,
}

\graphicspath{{./img/}}

\begin{document}

\title{Optimal transition states for polaron hopping transport without supercells}

\author{Vasilii Vasilchenko}
\thanks{Corresponding author: vasilii.vasilchenko@uclouvain.be}
\affiliation{European Theoretical Spectroscopy Facility, Institute of Condensed Matter and Nanosciences, Universit\'e catholique de Louvain, Rue de l'Observatoire 8, bte L07.03.01, B-1348 Louvain-la-Neuve, Belgium}
\author{Matteo Giantomassi}
\affiliation{European Theoretical Spectroscopy Facility, Institute of Condensed Matter and Nanosciences, Universit\'e catholique de Louvain, Rue de l'Observatoire 8, bte L07.03.01, B-1348 Louvain-la-Neuve, Belgium}
\author{Samuel Ponc\'e}
\affiliation{European Theoretical Spectroscopy Facility, Institute of Condensed Matter and Nanosciences, Universit\'e catholique de Louvain, Rue de l'Observatoire 8, bte L07.03.01, B-1348 Louvain-la-Neuve, Belgium}
\affiliation{WEL Research Institute, avenue Pasteur 6, 1300 Wavre, Belgium}
\author{Xavier Gonze}
\affiliation{European Theoretical Spectroscopy Facility, Institute of Condensed Matter and Nanosciences, Universit\'e catholique de Louvain, Rue de l'Observatoire 8, bte L07.03.01, B-1348 Louvain-la-Neuve, Belgium}

\date{\today}

\begin{abstract}
Polaron formation localizes charge carriers and drives a crossover from band-like to hopping transport in materials.
Hopping dynamics can be obtained from DFT supercell calculations of transition states, but these suffer from polaron self-interaction, spurious electrostatics, and poor scaling with polaron size.
We introduce a supercell-free framework for \textit{ab initio} polaron hopping transport based on the \textit{ab initio} polaron equations formalism, its variational formulation, and the string method.
The approach avoids explicit real-space DFT supercell calculations: it optimizes transition states directly in reciprocal space and provides the polaron configurations along the path, enabling the evaluation of adiabatic hopping rates and mobilities.
We apply the method to LiF and rutile TiO$_2$, revealing multi-step and anisotropic hopping mechanisms.
In rutile TiO$_2$, we compare the computed electron-polaron mobility with experiments and with alternative first-principles approaches.
Our results establish a scalable route to first-principles polaron-hopping dynamics in materials in which charge motion is governed by self-trapping.
\end{abstract}

\maketitle

\section{Introduction}
\label{sec:intro}

In materials, polarons form when particles interact with atomic vibrations and become dressed by a phonon cloud~\cite{Franchini2021}.
This process changes the nature of the particle and affects the optical and transport properties of the host medium.

Traditionally, polaron formation has been considered for electrons and holes in polarizable media with strong electron-phonon coupling (EPC), such as ionic crystals~\cite{Landau1933, Pekar1946a}.
In these systems, the charge carrier distorts the lattice, inducing a local polarization field.
In response, this polarization field interacts with the carrier and lowers its energy.
As a result, a polaron forms: the carrier propagates through the crystal accompanied by its self-induced phonon field.
More generally, polaronic effects have been observed and theoretically predicted in a broad range of materials, including non-polar~\cite{Vasilchenko2021, Zhang2025} and two-dimensional materials~\cite{Kang2018, Sio2023}, organic crystals~\cite{Zhugayevych2015, Chang2022}, and polymers~\cite{Bredas1985, Wu2025}.

Polarons underlie multiple physical phenomena, including the renormalization of electronic bands~\cite{Ciuchi2012} and the enhancement of effective masses~\cite{Yamada2022}.
Polaron formation may also lead to self-trapping, whereby a charge carrier becomes localized in the potential created by its self-induced polarization~\cite{Castner1957}.
The competition between band renormalization and self-trapping strongly influences the transport properties of a system.
Depending on the strength of the EPC, a system may undergo a crossover from band-like transport~\cite{Ponce2020} to polaron hopping~\cite{Natanzon2020}.
In the former case, charge carriers remain mobile polarons, whose transport properties are affected by the associated phonon field~\cite{Chang2022, Liu2026}.
In the latter scenario, polarons are self-trapped, and charge transfer occurs by hopping between localization sites.
Such transitions are thermally activated as they require the system to cross energy barriers between localized states.
As a result, in the hopping regime, the charge-carrier mobility is substantially reduced and may fall below 1~cm$^2$\,V$^{-1}$\,s$^{-1}$~\cite{Yagi1996, Zhang2007}.

Experimentally, polaronic signatures in solids can be detected and extracted using angle-resolved photoemission spectroscopy (ARPES)~\cite{Ciuchi2012, vanWaas2026}, electron paramagnetic resonance (EPR) spectroscopy~\cite{Yang2013}, scanning tunnelling microscopy (STM)~\cite{Setvin2014}, transport measurements~\cite{Crevecoeur1970}, and other techniques~\cite{Franchini2021}.
Recent advances in experimental methods and materials synthesis now allow a growing range of polaronic features to be probed, controlled, and analysed~\cite{RenedeCotret2022, Liu2023}, including hopping transport~\cite{Tyunina2023, MohamedJibri2023, Liu2026b}.

From the theoretical and computational perspective, various methods have been developed to address the problem of polaron formation: model Hamiltonians, supercell density-functional theory (DFT) calculations and effective first-principles approaches~\cite{Dai2025}.
The latter include diagrammatic Monte Carlo (DMC)~\cite{Luo2025}, canonical-transformation techniques~\cite{Lee2021, Luo2022, Robinson2025}, polaron Hamiltonian diagonalization~\cite{Sio2019a, Sio2019b}, and variational approaches~\cite{Vasilchenko2022, Baumgarten2026}.
Although built on techniques originally developed for model Hamiltonians, these effective methods treat electrons, phonons, and EPC \textit{ab initio}.
Thus, they provide a favourable trade-off between computational complexity and an \textit{ab initio} description of polarons in real materials.
Moreover, these approaches bypass key shortcomings of direct DFT polaron calculations~\cite{Dai2025}: the polaron self-interaction error, which requires dedicated self-interaction corrections (SIC)~\cite{Falletta2025}, spurious electrostatic interactions between periodic images~\cite{Komsa2012}, and unfavourable scaling with the polaron size.
Therefore, in recent years, such methods have received considerable attention and development effort, becoming important tools in computational polaron physics.
Their implementations are being developed in various \textit{ab initio} codes, such as \textsc{EPW}~\cite{Sio2019a, Sio2019b, Lee2023}, \textsc{Perturbo}~\cite{Luo2025, Zhou2021}, \textsc{ABINIT}~\cite{Verstraete2025, Vasilchenko2025}, and \textsc{Q-Chem}~\cite{Epifanovsky2021, Baumgarten2026}.

Beyond polaron formation, many of the aforementioned methods can also describe polaron transport.
In both the band-like and hopping regimes, \textit{ab initio} polaron charge transfer can be captured using the first-principles DMC method~\cite{Luo2025}.
This approach can be used to stochastically evaluate the Kubo current-current correlation function~\cite{Kubo1957} and compute transport quantities~\cite{Mishchenko2019}.
The purely adiabatic hopping of self-trapped polarons can also be described directly with DFT.
Indeed, the polaron potential energy surface (pPES) can be effectively sampled using DFT total-energy calculations, which also provide access to polaron forces.
These forces, in turn, are necessary for computing optimal transition paths~\cite{Sheppard2008}, which describe polaron hopping within the framework of transition-state theory~\cite{Marcus1993}.
However, the aforementioned limitations of DFT restrict its applicability.

Recently, the scalability of DFT-based polaron hopping calculations has been extended using machine-learning molecular dynamics (MLMD)~\cite{Birschitzky2025, Ahart2026}.
Another route towards scalable and efficient hopping calculations is provided by effective first-principles methods, which give access to polaron distortions.
However, so far, among these effective methods, hopping barriers have only been computed within the framework of \textit{ab initio} polaron equations introduced by Sio \textit{et al.}~\cite{Sio2019a, Sio2019b}, and only along linear transition paths~\cite{LafuenteBartolome2024, Dai2024c, Dai2026b}, thereby yielding upper bounds for the transition barriers.
Nonetheless, existing fully variational frameworks~\cite{Vasilchenko2022} provide access to polaron energetics and forces for arbitrary polaron configurations.
Thus, they can be extended to compute optimized minimum-energy paths (MEPs) for polaron transfer and to describe polaron hopping dynamics beyond the limitations of DFT.

Accordingly, in this work, we use the \textit{ab initio} polaron equations framework~\cite{Sio2019a, Sio2019b} and extend its variational formulation~\cite{Vasilchenko2022} to the calculation of MEPs.
Previously, we have demonstrated the capability of this formalism to describe self-trapped polarons in real materials~\cite{Vasilchenko2025}.
These polarons correspond to local minima on the pPES, defined by the variational polaron energy expression, which depends on both electronic and vibrational degrees of freedom.
The energy is expressed within the adiabatic strong-coupling approximation.
Within this approximation, the lattice distortion is treated as static, and quantum lattice fluctuations are neglected.
Under the strong-EPC assumption, the polaron charge is assumed to adjust instantaneously to the lattice polarization.
This separates the electronic and lattice degrees of freedom, while the variational nature of the formalism allows the corresponding electronic and phononic gradients to be evaluated.
The explicit treatment of gradients enables efficient and scalable optimization techniques for finding polaron solutions.
Moreover, it allows methods designed for computing MEPs, which connect these solutions on the pPES, to be implemented~\cite{Sheppard2008}.

For this purpose, we implement the simplified string method~\cite{E2007}.
For given initial and final polaron configurations, it yields the MEP corresponding to the hopping transition.
The MEPs are represented by a finite set of polaronic images, describing the evolution of the polaron energy, as well as its electronic and structural configuration, along the path.
Within transition-state theory, MEPs provide the key ingredients required to describe polaron dynamics~\cite{Marcus1993}.

We show how the developed formalism can be applied to the calculation of polaron hopping transport.
As representative cases, we consider two systems: LiF and rutile TiO$_2$.
LiF is a paradigmatic polaronic insulator, hosting both small hole and large electron polarons, and we use it to illustrate and benchmark our variational transport framework.
For the electron polaron in LiF, our calculations are primarily exploratory and serve to demonstrate the numerical applicability of the framework for polarons with extended distortions.
Also, being a wide-bandgap insulator, LiF primarily serves as a benchmark system, for which we do not attempt a direct comparison with experiment.
To address this limitation, we also consider rutile TiO$_2$, one of the most extensively studied polaronic semiconductors.
In this system, electron transport by small-polaron hopping has been predicted theoretically~\cite{Deskins2007}, and low electron mobilities, of the order of 1~cm$^2$\,V$^{-1}$\,s$^{-1}$ or below, have been measured experimentally~\cite{Yagi1996, Zhang2007}.

In both systems, we compute hopping polaron transfer rates and mobilities within transition-state theory.
We show that linear transition paths are insufficient to describe polaronic transition states.
We also show that hopping can be anisotropic, with certain directions being more energetically favourable than others.
For the hole polaron in LiF, which is triply degenerate, we show that hopping involves rotations between degenerate states as well as inter-site transitions.
For the electron polaron in rutile TiO$_2$, we show that the obtained hopping anisotropy is consistent with earlier DFT+$U$~\cite{Deskins2007} and MLMD calculations~\cite{Birschitzky2025}.
We also obtain reasonable agreement between our results and reported experimental measurements, and we discuss the differences with other theoretical methods.
We find that approaches which do not assume particle localization, such as the Boltzmann transport equation (BTE), may be insufficient to describe charge transfer in systems with bound polarons.
We also discuss the potential impact of non-adiabatic and weak-coupling effects~\cite{Lihm2026}, which are neglected in our formalism by construction.

Overall, the present framework showcases how effective first-principles approaches address polaron transport in the hopping regime.
Moreover, it could potentially be extended to more advanced all-coupling techniques~\cite{LafuenteBartolome2022, Baumgarten2026}.
Instead of relying on predetermined hopping paths or explicit supercells, the present method determines transition paths variationally.
This provides access to optimized activation barriers, attempt frequencies, and hopping mobilities within a single reciprocal-space polaron formalism.
Therefore, it provides a practical route for studying polaron-limited transport in polar semiconductors, oxides, halides, and other materials in which localized charge carriers determine functionality.

\section{Results}
\label{sec:results}

\subsection{Hole polaron dynamics in LiF}

\begin{figure}[t]
    \centering
    \includegraphics[width=0.4\linewidth]{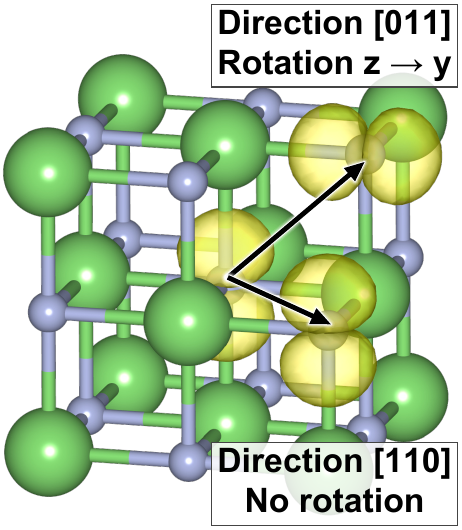}
    \caption{
    {\bfseries Hole polaron in LiF.}
    Examples of possible transitions for a $[001]$-oriented polaron solution.
    The polaron-density isosurface is set to $n^{\rm p}(\mathbf{r}) = 5\times10^{-3}$~\AA$^{-3}$.
    }
    \label{fig:hop-options}
\end{figure}

The hole polaron in LiF is localized on a fluorine atom, is triply degenerate, and exhibits $D_{4h}$ point-group symmetry~\cite{Vasilchenko2025}.
The polaron is strongly bound.
In the thermodynamic limit, its binding and localization (vertical ionization) energies are $E^{\rm pol} = -1.94~{\rm eV}$ and $\varepsilon^{\rm loc} = 4.68~{\rm eV}$, respectively~\cite{Vasilchenko2025}.
This polaron is small, with a characteristic length scale $a^{\rm p} = 2.26$~Bohr~\cite{Vasilchenko2025}.
Charge transfer can occur from a given localization site to one of the nearest-neighbour fluorine atoms.
Since each degenerate solution is oriented along one of the Cartesian axes, there are two distinct hopping directions: one with a component along the polaron axis and one lying in the plane perpendicular to it.
For instance, for a $[001]$-aligned solution, hopping may occur along the $[011]$ and $[110]$ directions, respectively.
Moreover, at each localization site, the polaron can switch between non-trivial degenerate configurations by rotating its principal axis.
As a result, there are six possible transitions that modify the $[001]$-aligned polaron configuration:
(i) on-site rotation to the $[100]$ or $[010]$ alignment, which are equivalent by symmetry;
(ii) two inter-site hopping processes along the $[110]$ direction: one preserves the $[001]$ alignment, while the other involves rotation to either the $[100]$ or $[010]$ symmetry-equivalent alignment;
(iii) three inter-site hopping processes along the $[011]$ direction: one preserves the $[001]$ alignment, while the other two involve rotation to the $[100]$ or $[010]$ alignments, which are symmetry-inequivalent for this direction.
Examples of these transitions are shown in Fig.~\ref{fig:hop-options}.

\begin{figure*}[t]
    \centering
    \includegraphics[width=1.\linewidth]{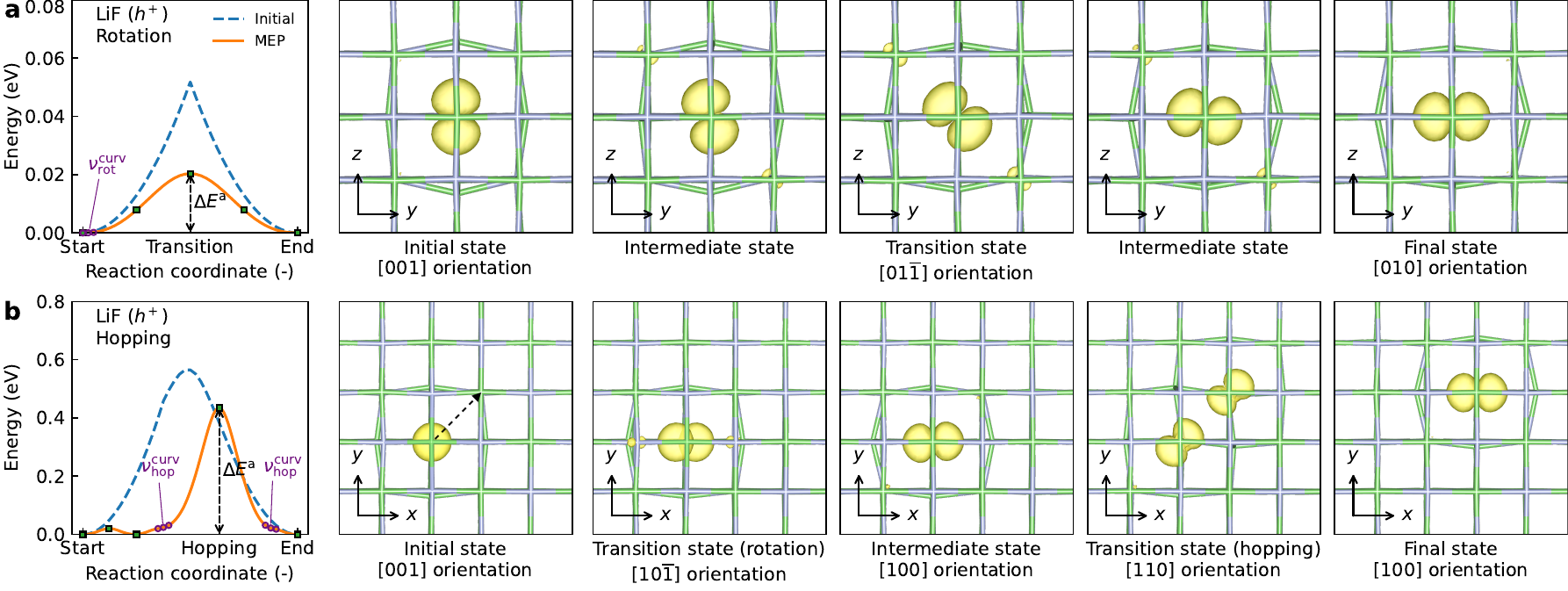}
    \caption{
    {\bfseries Hole polaron dynamics in LiF.}
    Rows {\bfseries a} and {\bfseries b} correspond to the polaron on-site $[001]\to[010]$ rotation and polaron hopping along the $[110]$ direction accompanied by on-site $[001]\to[100]$ rotation, respectively.
    The first panel in each row shows the initial linearly interpolated energy path and the optimized MEP.
    Other panels show the polaron in the initial, intermediate, transition, and final configurations, corresponding to the green markers on the MEP.
    The activation energy $\Delta E^{\rm a}$ is estimated from the initial state to the transition state.
    Purple open circles indicate the three images used to fit the MEP curvature and extract the attempt frequencies $\nu^{\rm curv}_{\rm rot}$ and $\nu^{\rm curv}_{\rm hop}$ for rotation and hopping (see Supplementary Note~S1).
    In the initial-state panel of {\bfseries b}, the dashed arrow indicates the $[110]$ hopping direction towards the destination fluorine site.
    The polaron-density isosurface is set to $n^{\rm p}(\mathbf{r}) = 5\times10^{-3}$~\AA$^{-3}$.
    }
    \label{fig:lif-polaron-hopping}
\end{figure*}

For an on-site rotation and an inter-site hopping event, Fig.~\ref{fig:lif-polaron-hopping} illustrates the initial linearly interpolated energy paths and the optimized MEPs.
The figure also shows the polaron configurations in the initial, transition, and final states along the MEPs.
The optimization lowers the energy barriers and samples the pPES more accurately, revealing multi-step adiabatic transition mechanisms.
For instance, the initial linearly interpolated path in Fig.~\ref{fig:lif-polaron-hopping}a corresponds to a diabatic transition dominated by the evolution of the atomic displacements.
Although the electronic coefficients are optimized at each image for a fixed distortion, the density remains close to one endpoint state before switching abruptly to the other near the cusp in the energy profile.
By contrast, path optimization yields a smooth adiabatic MEP along which both the polaron density and the atomic displacements evolve continuously, as illustrated in Fig.~\ref{fig:lif-polaron-hopping}a.
Inter-site hopping is more complex.
Irrespective of whether the polaron principal axis changes between lattice sites, such transitions require a specific polaron orientation before the hopping.
As a result, the energy paths may exhibit multiple barriers, corresponding to distinct transition states.
In particular, Fig.~\ref{fig:lif-polaron-hopping}b illustrates hopping along the $[110]$ direction accompanied by a change in the polaron orientation from the $[001]$ to the $[100]$ Cartesian axis.
This process involves two consecutive transition events: an on-site rotation and an inter-site hopping step.
The hopping transition state, shown in Fig.~\ref{fig:lif-polaron-hopping}b, is equivalent by symmetry for all five inter-site hopping processes listed above, two along $[110]$ and three along $[011]$.
This polaron configuration corresponds to the maximum of the MEP and has $D_{2h}$ point-group symmetry.
It corresponds to the two-centre $X^-_2$ polaron ($V_K$ centre) observed in alkali halides~\cite{Castner1957}, whereas the initial and final states correspond to one-centre $X^0$ polarons with $D_{4h}$ symmetry.
We note that, in several alkali halides, including LiF, the $X^-_2$ state has been observed experimentally by electron spin resonance as the stable configuration of the self-trapped hole~\cite{Castner1957}.
However, self-interaction-free approaches based on linear EPC~\cite{Sio2019a, Sio2019b, Lee2021, Robinson2025}, including the present one, consistently predict the $X^0$ polaron to be the stable lowest-energy solution in LiF.
The treatment of second-order EPC~\cite{Dai2026b}, together with anharmonic effects, could alter the relative stability of these solutions in favour of the $X^-_2$ polaron.

\begin{figure*}[t]
    \centering
    \includegraphics[width=0.8\linewidth]{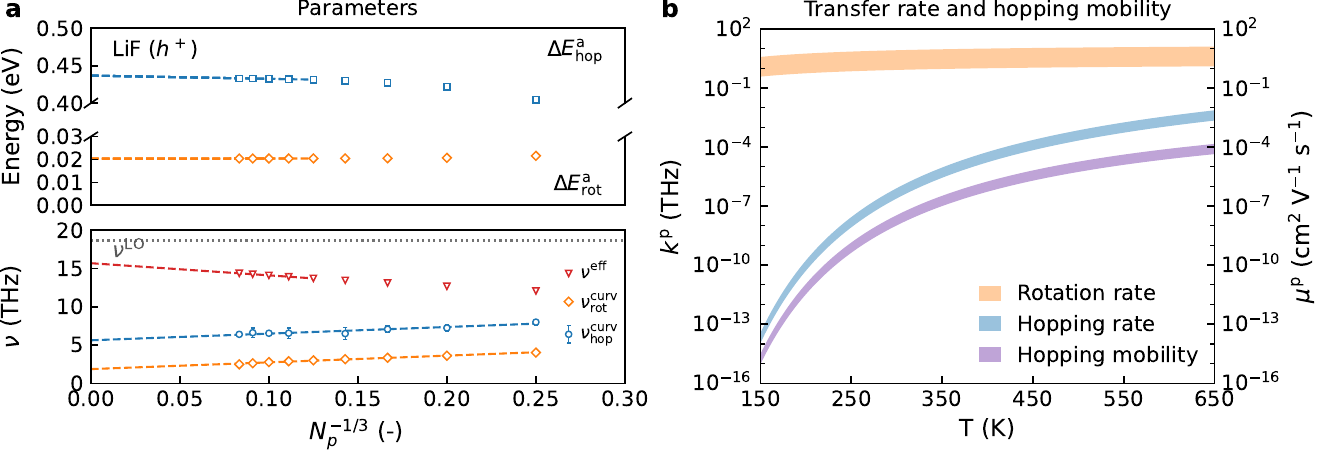}
    \caption{
    {\bfseries Charge-transfer parameters and mobility for the hole polaron in LiF.}
    {\bfseries a}~Activation energies $\Delta E_{\rm hop}^{\rm a}$ and $\Delta E_{\rm rot}^{\rm a}$ for hopping and rotation (top), and estimates of the attempt frequency (bottom): phonon-averaged $\nu^{\rm eff}$, curvature-based $\nu^{\rm curv}_{\rm hop}$ and $\nu^{\rm curv}_{\rm rot}$ for hopping and rotation, and zone-centre LO frequency $\nu^{\rm LO}$.
    These parameters are computed for $\mathbf{k/q}$-wavevector meshes with $N_p$ wavevectors, of increasing density from $4\times4\times4$ to $12\times12\times12$.
    Dashed lines indicate linear extrapolation to the thermodynamic limit.
    {\bfseries b}~Adiabatic transfer rates $k^{\rm p}$ (left axis) and hopping mobility $\mu^{\rm p}$ (right axis) as a function of temperature.
    The shaded areas span the values obtained with $\nu^{\rm curv}$ and $\nu^{\rm LO}$ as the attempt-frequency prefactor, which define the lower and upper limits, respectively.
    No mobility is associated with rotation, since it does not involve charge transfer between lattice sites.
    }
    \label{fig:lif-hole-electron-transport}
\end{figure*}

For both rotation and inter-site hopping, the extracted activation energies $\Delta E^{\rm a}$ and the estimated attempt frequencies $\nu^{\rm eff}$, $\nu^{\rm curv}$ and $\nu^{\rm LO}$, defined in the Methods section, are shown in Fig.~\ref{fig:lif-hole-electron-transport}a.
This figure also shows their extrapolation to the thermodynamic limit.
The phonon-averaged attempt frequency $\nu^{\rm eff}=15.69$~THz is identical for both processes because this quantity is evaluated at the initial states, which are instances of the same degenerate polaron configuration.
The process-dependent estimate based on the MEP curvatures yields lower values, $\nu^{\rm curv}_{\rm rot}=1.86$~THz and $\nu^{\rm curv}_{\rm hop}=5.62$~THz for rotation and hopping, respectively.
Unlike $\nu^{\rm eff}$ and $\nu^{\rm LO}$, which are properties of the initial polaron state or of the host crystal, $\nu^{\rm curv}$ reflects the softness of the pPES along the reaction coordinate of each process.
Since the hopping MEPs involve both rotation and hopping, $\nu^{\rm curv}_{\rm hop}$ is extracted by isolating the curvature of the hopping segment.
This segment is defined as the images around the hopping transition state whose energy exceeds the rotation barrier $\Delta E^{\rm a}_{\rm rot}$ obtained with the same $\mathbf{k/q}$-wavevector mesh.
The softness of the pPES is then estimated at the adjacent images below this threshold on each side of the transition
state, as indicated in Fig.~\ref{fig:lif-polaron-hopping}b and detailed in the Methods section and Supplementary Note~S1.

For the on-site rotation, we observe a significant reduction of the energy barrier with optimization: $\Delta E^{\rm a}_{\rm rot}=20$~meV for the optimized MEP, compared to $\Delta E^{\rm a}_{\rm rot}=50$~meV for the unoptimized path.
Moreover, this barrier is much smaller than the magnitude of the polaron binding energy, $|E^{\rm pol}|=1.94$~eV, and comparable to the thermal energy at room temperature, $k_B T \approx 26$~meV.
This indicates that transitions between degenerate configurations are accessible without the loss of polaron localization.
For the hopping processes, the barrier associated with the two-centre transition state is similar for all hopping events, with an average value of $\Delta E^{\rm a}_{\rm hop}=438\pm2$~meV.
The activation energy for translation is therefore more than 20 times larger than that for on-site rotation.
Thus, inter-site translation is the limiting step for polaron charge transfer in LiF.

The extracted parameters enable the evaluation of quantities describing polaron dynamics.
Specifically, we compute the adiabatic polaron transfer rates $k^{\rm p}$ and the mobility tensor $\boldsymbol \mu^{\rm p}$, which is isotropic.
The corresponding diffusion tensors of all the polarons considered in this work are derived in Supplementary Note~S2.
The mobility is not defined for on-site rotation, since no charge transfer between lattice sites occurs.
The temperature dependences of the transport quantities are shown in Fig.~\ref{fig:lif-hole-electron-transport}b.
The extrapolated transition-state theory parameters and room-temperature transfer rates are gathered in Table~\ref{tab:transfer-parameters}, with the components of the mobility tensor reported in Table~\ref{tab:tensor-mobilities}.
We note that the values of $k^{\rm p}$ and $\boldsymbol \mu^{\rm p}$ depend on the estimate of the attempt frequency $\nu$, which enters the definition of $k^{\rm p}$ as a prefactor, with $\nu \equiv \nu^{\rm curv}$ and $\nu \equiv \nu^{\rm LO}$ defining the lower and upper limits of the reported range.
Therefore, at room temperature, the rotation rate lies between 0.85 and 8.50~THz and increases with temperature.
This rate is of THz order, indicating a high transition probability between degenerate configurations localized on the same atom.
The corresponding rate for translation is only between 0.25 and 0.83~MHz.
This results in a small hole hopping mobility, $\mu^{\rm p}\sim10^{-8}$~cm$^2$\,V$^{-1}$\,s$^{-1}$, characteristic of a strongly bound small polaron.

\subsection{Electron polaron dynamics in LiF}

Compared to the hole polaron, the electron polaron in LiF is large and weakly bound.
In the thermodynamic limit, its binding and localization energies are $E^{\rm pol}=-230~{\rm meV}$ and $\varepsilon^{\rm loc}=-811~{\rm meV}$, respectively~\cite{Vasilchenko2025}.
The characteristic length scale is $a^{\rm p} = 11.81$~Bohr~\cite{Vasilchenko2025}.
Hence, weak-coupling effects, which are neglected in our variational formalism and transition-state theory, may be important for the electron polaron. 
Thus, the present analysis of the dynamics of this polaron is valid only within the strong-coupling and adiabatic approximations.
We also note that this analysis is exploratory and mainly demonstrates that the method can be numerically applied to polarons with extended distortions.
It does not establish that hopping governs electron polaron transport in this material, which may be better represented by methods like first-principles DMC~\cite{Luo2025}.

The electron polaron in LiF is localized on lithium atoms and is isotropic~\cite{Vasilchenko2025}.
Hence, all nearest-neighbour lithium sites are equivalent, and electron transfer occurs along the first-shell nearest-neighbour directions of the $\langle 110 \rangle$ family.
The corresponding hopping process, optimized MEP, extracted transition-state theory parameters, and calculated transfer rates and mobilities are presented in Supplementary Note~S3, and the diffusion tensor is derived in Supplementary Note~S2.
Tables~\ref{tab:transfer-parameters} and~\ref{tab:tensor-mobilities} summarize the parameters, extrapolated to the thermodynamic limit, and the computed room-temperature transport quantities.
The activation energy for the nearest-neighbour hopping is $\Delta E^{\rm a}=0.173$~meV.
This is approximately three orders of magnitude smaller than the binding energy and much smaller than the thermal energy $k_B T$ at all relevant temperatures.
The estimates for the attempt frequency yield $\nu^{\rm curv}=1.50$~THz and $\nu^{\rm eff}=13.08$~THz.
At room temperature, the adiabatic transfer rate and mobility lie in the ranges from 1.49 to 18.54~THz and from 0.095 to 1.183~cm$^2$\,V$^{-1}$\,s$^{-1}$, respectively, depending on the attempt-frequency prefactor.
This mobility is several orders of magnitude higher than that of the hole polaron, as expected for a large polaron with weaker binding.
However, although the curvature-based prefactor lowers the estimated transfer rate, the extremely small activation barrier does not establish a regime of rare hopping events, so that the assumption of uncorrelated adiabatic hops is questionable.
These results therefore illustrate the application of the framework to an extended polaron rather than establish a quantitatively validated description of its transport.

\subsection{Electron polaron dynamics in rutile TiO\textsubscript{2}}
\label{sec:tio2-polaron}

\begin{figure}[t]
    \centering
    \includegraphics[width=1.\linewidth]{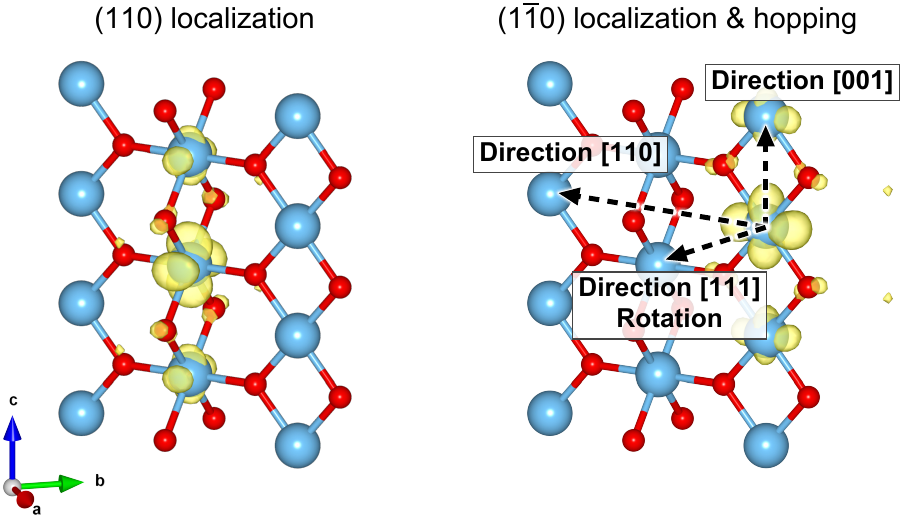}
    \caption{
    {\bfseries Electron polaron in rutile TiO$_2$.}
    Localization sites and hopping directions.
    The polaron-density isosurface is set to $n^{\rm p}(\mathbf{r}) = 5\times10^{-3}$~\AA$^{-3}$.
    }
    \label{fig:tio2-polaron-localization}
\end{figure}

\begin{figure*}[t]
    \centering
    \includegraphics[width=0.667\linewidth]{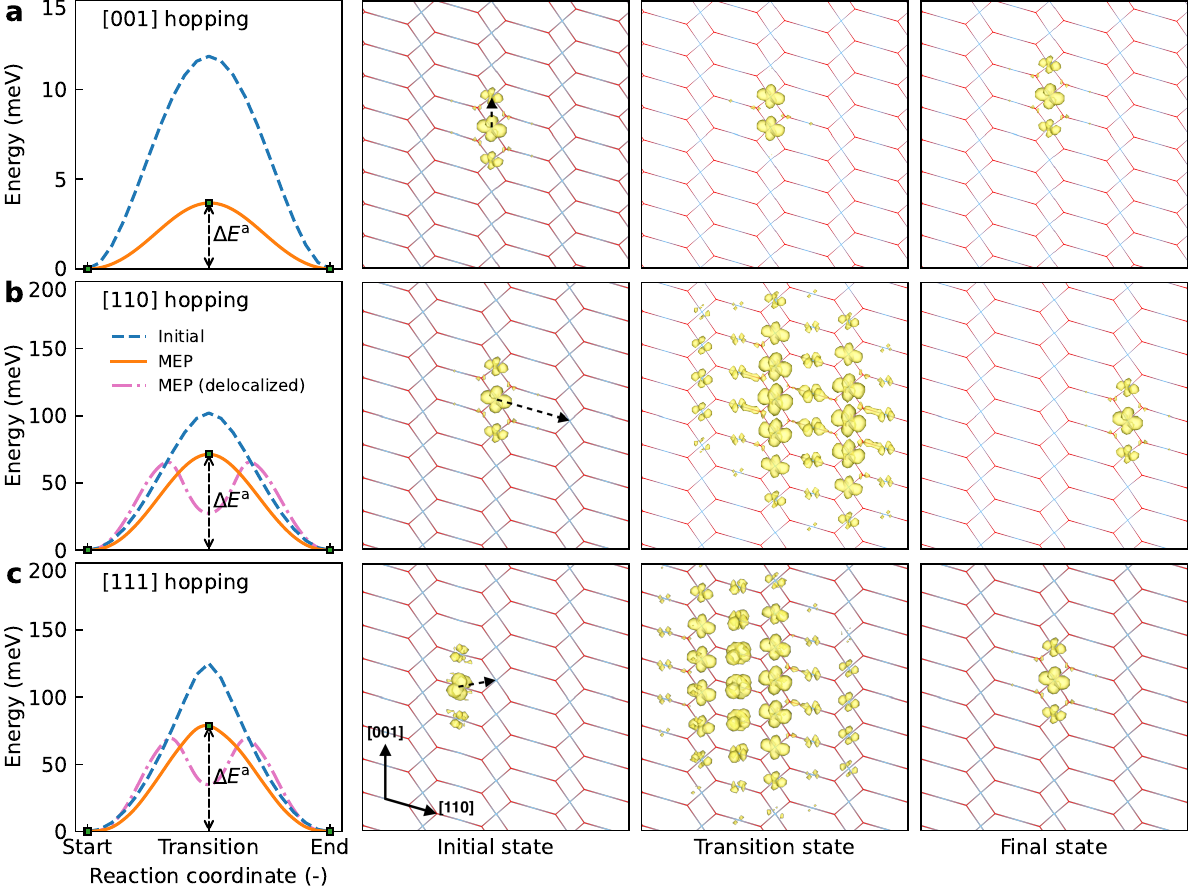}
    \caption{
    {\bfseries
    Electron polaron dynamics in rutile TiO$_2$.
    }
    Rows {\bfseries a}, {\bfseries b} and {\bfseries c} correspond to hopping along the $[001]$, $[110]$, and $[111]$ directions, respectively.
    Each row shows the initial linearly interpolated energy path, the optimized MEP, and the polaron in the initial, transition, and final configurations, corresponding to the green markers on the MEP.
    The activation energy $\Delta E^{\rm a}$ is estimated from the initial state to the transition state.
    In the initial-state panels, the dashed arrow points from the occupied Ti site towards the destination site of the hop.
    The configurations are shown in the $(1\overline{1}0)$ plane, which contains the $[001]$ and $[110]$ directions.
    For the $[110]$ and $[111]$ transitions, the pink curves denote MEPs corresponding to polaron delocalization at the transition state.
    For the transition states of the $[110]$ and $[111]$ hopping processes, the density isosurface is set to $n^{\rm p}(\mathbf{r}) = 5\times10^{-4}$~\AA$^{-3}$ to illustrate fine features.
    For all other states, the density isosurface is set to $n^{\rm p}(\mathbf{r}) = 5\times10^{-3}$~\AA$^{-3}$.
    Calculations correspond to anisotropic $\mathbf{k/q}$ meshes of increasing density: $n\times n\times 2n$, with $n = 6$, 9 and 11 for the $[001]$, $[110]$ and $[111]$ hopping.
    For the latter two processes, delocalized MEPs are shown for $n = 8$ and 9, respectively.
    }
    \label{fig:tio2-electron-hopping}
\end{figure*}

The electron polaron in rutile TiO$_2$ localizes on one of the Ti atoms occupying the $2a$ Wyckoff sites of the P$4_2$/mnm space group.
These localization sites are shown in Fig.~\ref{fig:tio2-polaron-localization}.
Each of the two Ti sites in the primitive cell has local orthorhombic point symmetry and nearest-neighbour Ti atoms lying in either the $(110)$ or $(1\overline{1}0)$ lattice plane.
Irrespective of the localization site, the polaron extends along the $[001]$ principal axis of the crystal.
There is no symmetry-equivalent degenerate polaron associated with the same site.

The extrapolated values of the binding energy $E^{\rm pol}$ and localization energy $\varepsilon^{\rm loc}$ indicate that the polaron is weakly bound, with $E^{\rm pol}=-0.095$~eV and $\varepsilon^{\rm loc}=-0.714$~eV, see Supplementary Note~S4.
This weak binding is consistent with hybrid-functional calculations showing that localized and delocalized excess electrons in rutile are close in energy~\cite{Janotti2013}.
Our findings also agree with the results of Ref.~\citenum{Dai2024c}, obtained with a similar reciprocal-space approach~\cite{Sio2019a, Sio2019b}: $E^{\rm pol}=-0.11$~eV and $\varepsilon^{\rm loc}=-0.76$~eV.
As discussed in Ref.~\citenum{Vasilchenko2025}, the difference arises mainly from the different interpolation schemes used for the electron-phonon matrix elements.
The computed characteristic length scale is $a^{\rm p}=4.95$~Bohr, indicating a small polaron.
For this polaron, we consider three inequivalent transitions, shown in Fig.~\ref{fig:tio2-polaron-localization}, which correspond to the hopping channels observed in MLMD simulations~\cite{Birschitzky2025}.
Without rotation, the polaron can transfer from one equivalent Ti site to another along the $[001]$ and $[110]$ directions.
In contrast, transfer along the $[111]$ direction requires a rotation that changes the plane in which the polaron density is primarily localized.

Electron-polaron hopping in TiO$_2$ is anisotropic, in agreement with Refs.~\citenum{Deskins2007} and \citenum{Birschitzky2025}.
For the three directions considered, we obtain MEPs with substantially different profiles, as shown in Fig.~\ref{fig:tio2-electron-hopping}.
The same figure also illustrates the corresponding hopping processes.
For the $[110]$ and $[111]$ directions, the MEPs depend on the $\mathbf{k/q}$-wavevector meshes.
Since these meshes define the corresponding Born-von Karman (BvK) supercells, they must be sufficiently large to accommodate the transition state.
All considered meshes are anisotropic, $n\times n\times 2n$, $6 \le n \le 11$, and are sufficient to host the initial and final polaron states.
However, the smallest meshes lead to charge transfer involving complete delocalization of the polaron at the transition state, followed by localization on another site.
These are finite-size effects, which
disappear at denser meshes, and such processes are discarded.

The MEPs in Fig.~\ref{fig:tio2-electron-hopping}b and~c illustrate this behaviour for hopping along the $[110]$ and $[111]$ directions, showing the qualitative difference between localized transfer and transfer involving delocalization.
We note that, for the $[110]$ and $[111]$ hopping processes, the meshes required to obtain localized transfer are at least $9\times9\times18$ and $11\times11\times22$, respectively.
In contrast, $[001]$ transfer remains localized for all considered meshes.
As shown below, it is also the most favourable transition, with the lowest activation energy.
The dominance of $[001]$ transfer is consistent with DFT+$U$ calculations~\cite{Deskins2007}.
It also agrees with MLMD calculations of polarons in TiO$_2$~\cite{Birschitzky2025}, which report transition distributions of $74.3$~\%, $25.5$~\%, and $0.2$~\% for the $[001]$, $[110]$, and $[111]$ directions, respectively.
However, MLMD finds $[111]$ transfer to be much less frequent than $[110]$ transfer, whereas we obtain similar activation energies for these two processes.
This difference may be related to the diabatic character of the $[111]$ transition reported in Ref.~\citenum{Birschitzky2025}, which cannot be captured within the present adiabatic framework.

\begin{figure*}[t]
    \centering
    \includegraphics[width=0.8\linewidth]{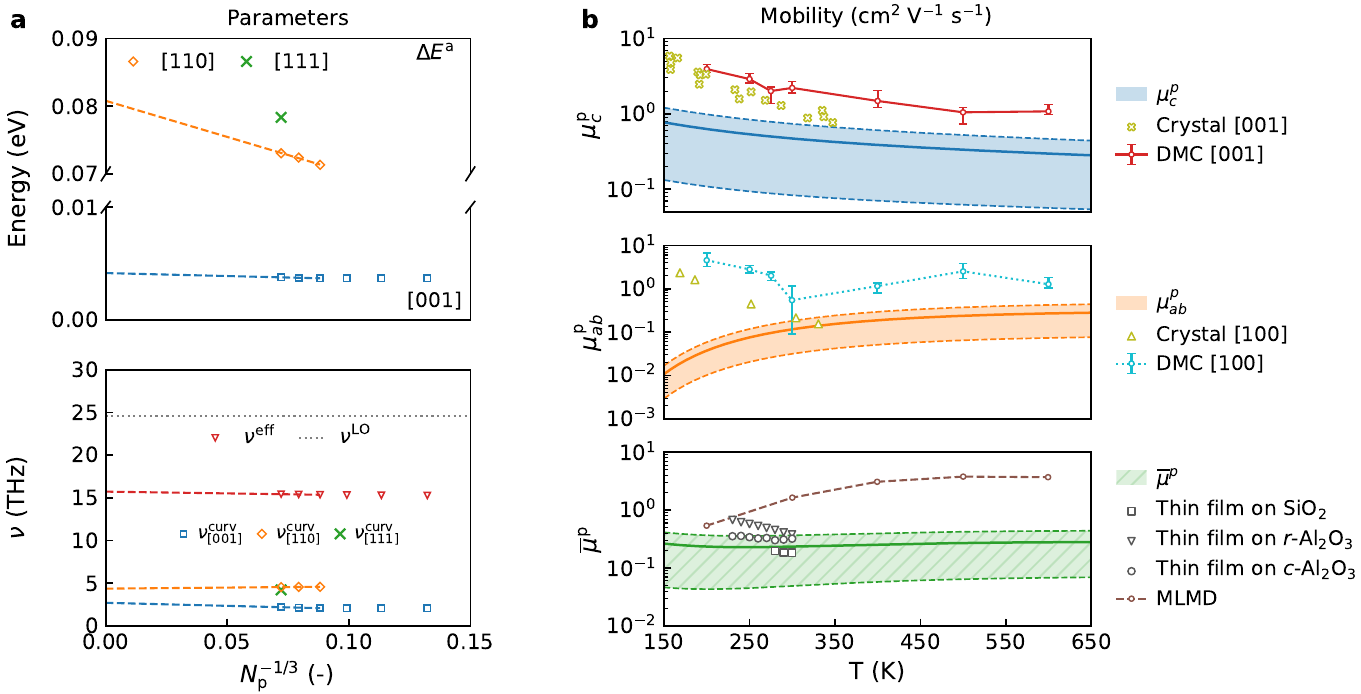}
    \caption{
    {\bfseries
    Charge-transfer parameters and mobility for the electron polaron in rutile TiO$_2$.
    }
    {\bfseries a}~Activation energies $\Delta E^{\rm a}$ (top) and estimates of the attempt frequency $\nu$ (bottom), computed for anisotropic $n\times n\times 2n$ $\mathbf{k/q}$ meshes with $6 \le n \le 11$.
    Localized transfer along $[110]$ and $[111]$ requires $n\ge9$ and $n\ge11$, respectively.
    Dashed lines show linear extrapolations to the thermodynamic limit, except for $[111]$, where only the $n=11$ values are used as the parameter estimates, since denser meshes with $n>11$ are computationally prohibitive.
    {\bfseries b}~Components $\mu^{\rm p}_{c}$ (top) and $\mu^{\rm p}_{ab}$ (middle) and average $\overline{\mu}^{\rm p}$ (bottom) of the hopping mobility tensor.
    The components are compared with direction-resolved measurements in oxygen-deficient single crystals~\cite{Yagi1996} and with first-principles DMC results~\cite{Luo2025}; the average is compared with direction-averaged measurements on Nb-doped thin films~\cite{Zhang2007} and with the MLMD mobility~\cite{Birschitzky2025}.
    Shading spans the mobility values obtained using $\nu^{\rm curv}$ and $\nu^{\rm LO}$ as the attempt-frequency prefactor, and the solid curves within the shaded regions indicate the results obtained with $\nu^{\rm eff}$.
    }
    \label{fig:tio2-electron-transport}
\end{figure*}

The extracted transition-state-theory parameters are shown in Fig.~\ref{fig:tio2-electron-transport}a and gathered in Table~\ref{tab:transfer-parameters}.
For hopping along the $[111]$ direction, the parameters are estimated using only the largest considered mesh.
Smaller meshes yield delocalized transitions, and are therefore discarded.
Denser meshes are computationally prohibitive because of the memory needed to precompute the electron-phonon matrix elements for the polaron~\cite{Verstraete2025}.
We therefore do not extrapolate the parameters for this transition to the thermodynamic limit.
In principle, this should not qualitatively affect the results.
We note, however, that for the $[110]$ transition, the extrapolation increases the activation energy by about 8~meV, i.e., by approximately 10~\%, with respect to the value obtained with the $11\times11\times22$ mesh.
A similar correction may apply to the $[111]$ transition, whose activation energy may therefore be underestimated.

The initial linearly interpolated energy barriers agree with those obtained using a similar reciprocal-space approach~\cite{Sio2019a, Sio2019b}, but without energy-path optimization~\cite{Dai2024c}.
For hopping along the $[001]$ and $[111]$ directions, our initial activation energies are $\Delta E^{\rm a, in}=12$~meV and $\Delta E^{\rm a,in}=124$~meV, respectively, in agreement with Ref.~\citenum{Dai2024c}, which reports $\Delta E^{\rm a, in}=13$~meV and $\Delta E^{\rm a, in}=133$~meV for the same transitions.
However, after energy-path optimization, these barriers are reduced to $\Delta E^{\rm a}=4$~meV and $\Delta E^{\rm a}=78$~meV, respectively.
This highlights the importance of energy-path optimization, especially for the most favourable $[001]$ transition, for which the optimized barrier is only one third of the initial value.
Supplementary Note~S5 provides an additional comparison of these direction-resolved barriers with other theoretical methods.
In particular, the seminal DFT+$U$ study of Deskins and Dupuis~\cite{Deskins2007} yields substantially larger barriers of 90 and 310~meV along $[001]$ and $[111]$, respectively.
These differences may reflect the use of a large Hubbard parameter ($U^{\rm eff}=10$~eV), linearly interpolated transition paths, and the nonadiabatic treatment of $[111]$ transfer in that work.

\begin{table*}[t]
    \setlength{\tabcolsep}{3pt}
    \centering
    \caption{
    Transition-state theory parameters and transfer rates for the polaron transfer considered in this work.
    The parameters include activation energies $\Delta E^{\rm a}$ and estimates of the attempt frequency $\nu^{\rm eff}$, $\nu^{\rm LO}$ and $\nu^{\rm curv}$.
    The adiabatic polaron transfer rates $k^{\rm p}$ are given at room temperature.
    For the LiF hole polaron, hopping parameters are averaged over the five possible transitions.
    }
    \label{tab:transfer-parameters}
  \begin{tabular*}{0.8\textwidth}{@{\extracolsep{\fill}}lrrrrrrr@{}}
    \hline\hline
    & \multirow{2}{*}{$\Delta E^{\rm a}$ (meV)}  & \multicolumn{3}{c}{Attempt frequencies (THz)} & \multicolumn{3}{c}{Transfer rates at 300~K (THz)} \\
    Process &  & \makebox[1.3cm][r]{$\nu^{\rm curv}$}
      & \makebox[1.3cm][r]{$\nu^{\rm eff}$} & \makebox[1.3cm][r]{$\nu^{\rm LO}$}
      & $k^{\rm p,curv}$ & $k^{\rm p,eff}$ & $k^{\rm p,LO}$ \\
    \hline
    \multicolumn{8}{@{}l@{}}{LiF ($h^+$)} \\
    Rotation & 20.32 & 1.86 & \multirow{2}{*}{15.69} & \multirow{2}{*}{18.66} & 0.85 & 7.15 & 8.50 \\
    Hop & 437.76 & 5.62 & & & $2.49\times10^{-7}$ & $6.95\times10^{-7}$ & $8.26\times10^{-7}$ \\
    \hline
    \multicolumn{8}{@{}l@{}}{LiF ($e^-$)} \\
    Hop & 0.17 & 1.50 & 13.08 & 18.66 & 1.49 & 13.00 & 18.54 \\
    \hline
    \multicolumn{8}{@{}l@{}}{TiO$_2$ ($e^-$)} \\
    $[001]$ hop & 3.86 & 2.70 & \multirow{3}{*}{15.72} & \multirow{3}{*}{24.61} & 2.33 & 13.54 & 21.20 \\
    $[110]$ hop & 80.86 & 4.37 & & & 0.19 & 0.69 & 1.08 \\
    $[111]$ hop & 78.01 & 4.21 & & & 0.21 & 0.77 & 1.20 \\
    \hline\hline
  \end{tabular*}

\end{table*}

\begin{table}[t]
  \centering
  \setlength{\tabcolsep}{3pt}
  \caption{
  Components of the mobility tensor $\boldsymbol\mu^{\rm p}$ for the polaron transfer considered in this work.
  Superscripts indicate the values obtained with different estimates for the attempt frequency: $\nu^{\rm eff}$, $\nu^{\rm LO}$ and $\nu^{\rm curv}$.
  For TiO$_2$, the tensor average is $\overline\mu^{\rm p}=\operatorname{Tr} (\boldsymbol\mu^{\rm p})/3$.
  }
  \label{tab:tensor-mobilities}
  \begin{tabular*}{\columnwidth}{@{\extracolsep{\fill}}lrrr@{}}
    \hline\hline
    & \multicolumn{3}{c}{Mobility at 300~K (cm$^2$\,V$^{-1}$\,s$^{-1}$)} \\
    Component & $\mu^{\rm p,curv}$ & $\mu^{\rm p,eff}$ & $\mu^{\rm p,LO}$ \\
    \hline
    \multicolumn{4}{@{}l@{}}{LiF ($h^+$)} \\
    Isotropic & $1.06\times10^{-8}$ & $2.95\times10^{-8}$ & $3.51\times10^{-8}$ \\
    \hline
    \multicolumn{4}{@{}l@{}}{LiF ($e^-$)} \\
    Isotropic & 0.0949 & 0.829 & 1.183 \\
    \hline
    \multicolumn{4}{@{}l@{}}{TiO$_2$ ($e^-$)} \\
    $ab$ plane & 0.0317 & 0.116 & 0.182 \\
    $c$ axis & 0.0835 & 0.471 & 0.738 \\
    Average & 0.0490 & 0.235 & 0.367 \\
    \hline\hline
  \end{tabular*}

\end{table}

The phonon-averaged estimate of the attempt frequency yields $\nu^{\rm eff} = 15.72$~THz and is independent of the hopping process.
It is lower than the zone-centre LO mode frequency $\nu^{\rm LO} = 24.61$~THz, which has been used as an estimate for $\nu$ in previous studies~\cite{Deskins2007, Dai2024c}.
However, the curvature-based estimate yields the lowest attempt frequencies, with a different value for each process: $\nu^{\rm curv}_{[001]} = 2.70$~THz, $\nu^{\rm curv}_{[110]} = 4.37$~THz, and $\nu^{\rm curv}_{[111]} = 4.21$~THz.
The curvature-based estimate accounts for the local energy variation along each optimized MEP and yields process-dependent attempt frequencies.
However, for the low-barrier $[001]$ transition, the activation energy is smaller than the thermal energy at room temperature, and the calculated transfer rate approaches the attempt frequency.
The assumption of well-separated, thermally activated hopping events is therefore not guaranteed, and the neglect of correlations between successive hops, including back-hopping, is questionable in this regime.
The corresponding mobility should therefore be interpreted as an estimate within the present adiabatic framework.
Nonetheless, as for polarons in LiF, we use $\nu^{\rm curv}$ and $\nu^{\rm LO}$ to obtain lower and upper estimates of the attempt frequency and mobility tensor.

The computed room-temperature transfer rates and mobility tensor components are summarized in Tables~\ref{tab:transfer-parameters} and~\ref{tab:tensor-mobilities}.
The mobility is anisotropic, and its temperature dependence is shown in Fig.~\ref{fig:tio2-electron-transport}b.
The corresponding diffusion tensor is derived in Supplementary Note~S2.
The $[001]$ transfer is the most favourable, with a room-temperature transfer rate between 2.33 and 21.20~THz, depending on the estimate of the attempt frequency $\nu$, which is at least one order of magnitude higher than those of the other transitions.
At room temperature, the $c$ component of the mobility tensor lies between 0.084 and 0.738~cm$^2$\,V$^{-1}$\,s$^{-1}$, depending on the estimate of $\nu$, which is typical of polaron hopping.
The averaged mobility $\overline\mu^{\rm p}=\operatorname{Tr} (\boldsymbol\mu^{\rm p})/3$ lies between 0.049 and 0.367~cm$^2$\,V$^{-1}$\,s$^{-1}$.

The comparison of the calculated hopping mobility tensor components $\mu^{\rm p}_{ab}$ and $\mu^{\rm p}_{c}$ and of the average $\overline\mu^{\rm p}$ with experimental data is shown in Fig.~\ref{fig:tio2-electron-transport}b.
The reference experiments are based on resistivity and carrier-concentration measurements in rutile TiO$_2$, using Nb-doped thin films grown on various substrates~\cite{Zhang2007} and oxygen-deficient single crystals~\cite{Yagi1996}.
We observe that $\overline\mu^{\rm p}$ is in reasonable agreement with the direction-averaged mobility measured in Nb-doped thin films.
The $\mu^{\rm p}_{c}$ component is lower than the direction-resolved mobility measured along $[001]$ in oxygen-deficient single crystals, but reproduces its decrease with increasing temperature.
The $\mu^{\rm p}_{ab}$ component increases with temperature, whereas the measured mobility along $[100]$ decreases, so that the two become comparable only above 300~K.
We note, however, that our calculations are performed for an ideal bulk TiO$_2$ crystal with a single polaron in the dilute limit.
This is a considerable simplification of the experimental samples, where oxygen vacancies, Nb dopants, carrier concentration, and surfaces or interfaces can affect polaron localization and mobility.

We also compare the computed hopping mobilities with reference data obtained using other first-principles approaches to polaron dynamics.
Figure~\ref{fig:tio2-electron-transport}b shows the comparison with MLMD~\cite{Birschitzky2025} and first-principles DMC~\cite{Luo2025} results.
The mobility tensor component along the $c$-direction exhibits the same decreasing temperature dependence in our calculations and in DMC, although the latter yields larger values.
For the in-plane component $\mu^{\rm p}_{ab}$, our mobilities increase with temperature, consistent with the trend of the corresponding DMC mobility at $T > 300$~K.
The discrepancies with DMC may arise because DMC is an all-coupling approach, whereas our formalism is restricted to the strong-coupling limit.
Although our results should be directly comparable to DFT-based supercell calculations~\cite{Dai2026}, the direction-averaged $\overline\mu^{\rm p}$ differs from the MLMD mobility, showing a different temperature dependence.
This difference may arise because MLMD is trained on self-consistent DFT data and is not restricted to harmonic lattice dynamics or linear EPC.
A recent work has shown that nonlinear EPC can modify both polaron and hopping-barrier energetics, yielding results closer to those obtained from DFT supercell calculations, even if linearly interpolated transition paths are used~\cite{Dai2026b}.
Including such effects in our formalism may therefore improve its agreement with MLMD.
Moreover, the direction-averaged MLMD mobility is extracted from the mean-squared displacement of finite-temperature trajectories sampling multiple correlated hopping pathways, whereas our mobility tensor combines uncorrelated hops along the individual channels.
Finally, the activation barriers and attempt frequencies entering our mobilities are evaluated in the $T=0$~K limit and may change as the pPES and MEPs evolve with temperature.
By contrast, MLMD samples finite-temperature trajectories in the canonical ensemble, while DMC evaluates the finite-temperature current-current correlation function.

We also compute the phonon-limited mobility using the iterative Boltzmann transport equation (iBTE) framework~\cite{Ponce2020}, as implemented in the \textsc{ABINIT} software package~\cite{Brunin2020a, Brunin2020b, Claes2022}.
The resulting values, shown in Supplementary Note~S6, overestimate both the hopping mobilities and the experimental values, the latter by one to two orders of magnitude.
Part of this discrepancy may arise from impurity and defect scattering in the doped and reduced samples, which is not included in our iBTE calculations.
However, the magnitude of the overestimation reflects the fact that the iBTE does not assume carrier localization and describes band-like transport, making it insufficient for semiconductors with self-trapped polarons.
Thus, in such systems, charge transfer is better described within the hopping picture, as done in this work.

\section{Discussion}
\label{sec:discussion}

The central result of this work is a scalable first-principles route to polaron hopping transport without explicit DFT supercell calculations.
By combining variational polaron equations with the simplified string method, we obtain optimized minimum-energy paths between polaronic states and extract the corresponding transition-state-theory parameters.
This extends effective first-principles polaron methods from the calculation of localized states to the calculation of hopping paths, rates, and mobilities.
We apply the method to polarons in LiF and the electron polaron in rutile TiO$_2$, demonstrating its capability to describe polaron hopping in realistic materials.

We show that the formalism can capture anisotropic hopping processes involving multiple intermediate configurations.
For all cases, we compute minimum-energy paths between polaronic states and estimate the corresponding adiabatic hopping mobilities.
The required adiabatic transport parameters are readily extracted from the energetics and polaron configurations along the MEPs.
Our framework also highlights the importance of energy-path optimization: barriers obtained along linearly interpolated paths are upper bounds and lead to underestimated hopping mobilities.
Moreover, the optimization can reveal complex charge-transfer mechanisms involving multiple intermediate configurations and transition states, as shown for the LiF hole polaron.

For the electron polaron in rutile TiO$_2$, we compare the hopping mobility with previously reported experimental data and results from state-of-the-art first-principles methods, finding qualitative agreement.
We hypothesize that quantitative differences arise primarily from the approximations included in the method: an ideal bulk crystal, harmonic phonons, linear EPC, and neglect of non-adiabatic and finite-temperature changes to the pPES.
We further show that the widely used Boltzmann transport equation framework overestimates the measured carrier mobility by one to two orders of magnitude, largely because it assumes band-like transport and does not account for carrier localization.
Therefore, for semiconductors hosting small self-trapped polarons, our methodology may provide a more appropriate description of charge transfer.

Broadly, the present formalism provides a practical step towards bridging two separate descriptions of charge transport: band-like electron-phonon transport and localized hopping of self-trapped carriers.
For materials in which the relevant carriers are localized, the method offers a way to compute direction-resolved hopping channels, compare competing transitions, and identify rate-limiting processes.
It can also help to assess when conventional Boltzmann transport becomes inappropriate.
Because the approach avoids explicit DFT supercell calculations, it is well suited for systematic studies of polaron-limited transport in realistic materials.
Future extensions beyond the adiabatic strong-coupling limit, including non-adiabatic, weak-coupling, anharmonic, and finite-temperature effects, would further broaden its applicability.
We also expect that the inclusion of correlations between hopping processes and a more general description of the hopping attempt frequencies would lead to a more accurate description of polaron hopping.

In summary, we have shown that transition states and minimum-energy paths for polaron hopping can be obtained directly from the variational formulation of the \textit{ab initio} polaron equations.
In contrast to real-space DFT calculations of total energies and forces, this approach describes the polaron potential-energy surface in reciprocal space, uses only primitive-cell inputs, and eliminates polaron self-interaction by construction.
It thus provides a computationally efficient alternative to DFT-based methods for calculating polaron hopping pathways.
Path optimization lowers hopping barriers relative to linearly interpolated paths, by a factor of three for the dominant $[001]$ transfer in rutile TiO$_2$, and reveals multistep mechanisms, such as rotation-assisted hole-polaron hopping in LiF.
The resulting adiabatic mobilities are consistent with the localized character of these carriers and, for TiO$_2$, with the measured mobility scale and anisotropy.
These results establish the variational polaron equations framework as a practical tool for identifying rate-limiting hopping channels and estimating carrier mobilities in systems where the adiabatic hopping description is applicable.

\section{Methods}
\label{sec:methods}

In this work, we investigate adiabatic polaron hopping transport within the variational formulation~\cite{Vasilchenko2022} of the \textit{ab initio} polaron equations framework~\cite{Sio2019a, Sio2019b}.
In particular, we extend the existing methodology, implemented in the \textsc{ABINIT} software package~\cite{Verstraete2025, Vasilchenko2025}, to enable the calculation of MEPs on the pPES using the simplified string method~\cite{E2007}.

In this section, we first provide the computational details used to obtain the input parameters for the materials considered in this work, LiF and rutile TiO$_2$.
We then recall the variational polaron equations framework and its range of applicability.
We also discuss its main features, namely the explicit variational expression for the polaron binding energy and the corresponding gradients.
Next, we show how these features enable sampling of the pPES and optimization of MEPs using the string method.
Then, we compare the computational cost of this approach with that of explicit DFT supercell calculations.
Finally, we describe transition-state theory for adiabatic transitions, as treated within the developed approach.

AI-assisted tools based on GPT-6.1 Sol and GPT-6 Astra  and Claude Opus 5.5  were used for proofreading, comparison of results with existing literature, and exploratory data analysis.
The authors have reviewed and verified all outputs and take full responsibility for the content of the work.

\subsection{Computational details}
\label{sec:comp-details}

Polarons are obtained using the variational polaron equations framework, as implemented in the \textsc{ABINIT} software package~\cite{Verstraete2025}.
MEPs are computed using our implementation of the simplified string method within this framework.
The input parameters are computed with \textsc{ABINIT}, using the PBE~\cite{Perdew1996b} exchange-correlation functional for LiF and the LDA for rutile TiO$_2$.
The LDA is used for rutile TiO$_2$ since PBE yields imaginary phonon frequencies in this system.
We use version 0.5 of the norm-conserving pseudopotential set~\cite{Hamann2013} from the \textsc{PseudoDojo} project~\cite{vanSetten2018}.
The atomic positions and lattice parameters of each system are relaxed until the maximum absolute force falls below $10^{-6}$~Ha/Bohr.
Supplementary Note~S7 reports the relaxed structural parameters and plane-wave energy cutoffs.

All the input quantities are computed in the primitive cell.
The KS electronic eigenvalues $\varepsilon_{n\mathbf{k}}$ and wavefunctions $\psi_{n\mathbf{k}}$ are computed self-consistently on a coarse $\mathbf{k}$-wavevector mesh.
The resulting electronic density is then used in non-self-consistent (non-SCF) calculations to obtain these quantities on dense $\mathbf{k}$-wavevector meshes of arbitrary size.
Phonon frequencies $\omega_{\mathbf{q}\nu}$, eigenmodes $e_{\kappa\alpha\nu}(\mathbf{q})$, and first-order derivatives of the KS potential $\partial_{\kappa\alpha\mathbf{q}} v^\mathrm{KS}$ are computed with density-functional perturbation theory (DFPT) on a coarse $\mathbf{q}$-wavevector mesh and subsequently interpolated onto dense $\mathbf{q}$-wavevector meshes~\cite{Eiguren2008, Brunin2020a, Brunin2020b}.
The electron-phonon matrix elements $g_{mn\nu}(\mathbf{k}, \mathbf{q})$ are then computed on the dense $\mathbf{k/q}$-wavevector meshes by Fourier interpolation of the short-range part of the scattering potentials, while the long-range dipole and quadrupole contributions are treated analytically~\cite{Verdi2015, Brunin2020a, Brunin2020b, Jhalani2020}.

In the latter step, to reduce memory requirements, the ratio between the FFT box and the plane-wave sphere is set to \texttt{boxcutmin}~$=1.1$.
The DFPT scattering potentials are therefore Fourier interpolated onto a coarser real-space FFT mesh, while all other calculations use the default value of \texttt{boxcutmin}~$=2$.
However, a recently identified bug in \textsc{ABINIT}, present in the version used in this work~\cite{matcloud_archive}, incorrectly handles some G-vectors when DFPT potentials are rotated to reconstruct $\mathbf{q}$-points by symmetry.
Its effect is negligible for \texttt{boxcutmin}~$\geq 2$, but it can affect electron-phonon calculations for FCC and hexagonal lattices when \texttt{boxcutmin}~$<2$.
Nonetheless, the results reported here with \texttt{boxcutmin}~$=1.1$ have been validated against those obtained with the default value on smaller meshes, and no discrepancies were observed.


Supplementary Note~S7 reports the coarse $\mathbf{k}$- and $\mathbf{q}$-wavevector meshes used to compute the electronic and phonon parameters, and the largest dense $\mathbf{k/q}$-wavevector meshes used in the polaron calculations for each system.
It also lists the $\mathbf{k/q}$-wavevector meshes required for computing the dielectric tensor, Born effective charges, and dynamical quadrupoles required for the long-range treatment of EPC.

\subsection{Variational polaron equations}

In the variational polaron equations framework~\cite{Vasilchenko2025}, the \textit{ab initio} polaron binding energy is defined in reciprocal space as a variational expression,
\begin{multline}\label{eq:epol}
    E^{\rm pol}[\boldsymbol{A}, \boldsymbol{B}] =
    \\
    \frac{\sigma}{N_p} \sum_{n\mathbf{k}} |A_{n\mathbf{k}}|^2 (\varepsilon_{n\mathbf{k}} - \varepsilon^{\rm loc}) + \sigma\varepsilon^{\rm loc}
    +\frac{1}{N_p} \sum_{\mathbf{q}\nu} |B_{\mathbf{q}\nu}|^2\omega_{\mathbf{q}\nu} \\
    -\frac{1}{N_p^2} \sum_{\substack{mn\mathbf{k} \\ \mathbf{q}\nu}}
    A^*_{m\mathbf{k + q}}
    B^*_{\mathbf{q}\nu}
    g_{mn\nu}(\mathbf{k}, \mathbf{q})
    A_{n\mathbf{k}}
    +
    \text{(c.c.)},
\end{multline}
where $\sigma =\pm1$ corresponds to electron addition/removal.
In the energy and displacement expressions below, we use atomic units ($\hbar=1$), so that phonon frequencies are expressed in energy units.
The energy depends on the electron and phonon parameters of the material: electronic energies $\varepsilon_{n\mathbf{k}}$, phonon frequencies $\omega_{\mathbf{q}\nu}$, and electron-phonon matrix elements $g_{mn\nu}(\mathbf{k},\mathbf{q})$.
The $\mathbf{k/q}$-wavevector meshes, each with $N_p$ points, correspond to a BvK supercell consisting of $N_p$ unit cells hosting the polaron.
Note that there are no explicit real-space calculations in the formalism, so that the direct use of supercells is avoided.

The variational subspace is defined by the sets of electronic and phonon coefficients $\boldsymbol{A} = \{ A_{n\mathbf{k}} \}$ and $\boldsymbol{B} = \{ B_{\mathbf{q}\nu} \}$.
The electronic coefficients originate from the expansion of the polaron wavefunction $\phi^{\rm p} (\mathbf{r})$ in the basis of Kohn-Sham (KS) states $\psi_{n\mathbf{k}} (\mathbf{r})$,
\begin{equation}\label{eq:phi-p}
    \phi^{\rm p} (\mathbf{r}) = \frac{1}{\sqrt{N_p}} \sum_{n\mathbf{k}} A_{n\mathbf{k}} \psi_{n\mathbf{k}}(\mathbf{r}).
\end{equation}
The normalization of the polaron wavefunction, $\langle \phi^{\rm p} |\phi^{\rm p} \rangle = 1$, is enforced in Eq.~\eqref{eq:epol} through the Lagrange multiplier $\varepsilon^{\rm loc}$.
This quantity represents the polaron localization (vertical ionization) energy.
The phonon coefficients represent the deformation field, being the components of the Fourier transform of the polaron-induced displacements $\Delta \boldsymbol{\tau} = \{ \Delta \tau_{\kappa\alpha p} \}$ in the phonon-momentum subspace,
\begin{equation}\label{eq:delta-tau}
    \Delta \tau_{\kappa \alpha p}
    =
    -\sigma \frac{2}{N_p}
    \sum_{\mathbf{q}\nu}
    B^*_{\mathbf{q}\nu}
    \frac{e_{\kappa\alpha\nu}(\mathbf{q})}
    {\sqrt{2M_\kappa \omega_{\mathbf{q}\nu}}}
    e^{i\mathbf{q}\cdot\mathbf{R}_p}.
\end{equation}
Here, for an atom $\kappa$ with mass $M_\kappa$ in unit cell $p$, $e_{\kappa\alpha\nu}(\mathbf{q})$ denotes the Cartesian component $\alpha$ of the phonon eigenmode.

Equations~\eqref{eq:epol}-\eqref{eq:delta-tau} describe self-trapped polarons in the strong-coupling adiabatic (Landau-Pekar) approximation~\cite{Pekar1946a}.
Within this approximation, lattice fluctuations are neglected, and the electronic charge density is assumed to adjust instantaneously to the lattice polarization due to strong electron-phonon interactions.
Accordingly, the polaron wavefunction $\phi^{\rm p}(\mathbf{r};\Delta \boldsymbol{\tau})$ depends parametrically on the structural distortion $\Delta \boldsymbol{\tau}$.
Moreover, this dependence is treated to lowest order in $\Delta \boldsymbol{\tau}$, corresponding to a frozen deformation that defines the adiabatic self-trapping potential.
The resulting polaron binding energy determines the adiabatic pPES governing the structural distortions.

Constrained optimization of the binding energy, $\min_{\langle \phi^{\rm p}|\phi^{\rm p} \rangle = 1} E^{\rm pol} [\boldsymbol{A}, \boldsymbol{B]}$, as well as sampling of the pPES, is facilitated by the explicit treatment of the electronic and phonon gradients:
\begin{align}
    \nabla_{A_{n\mathbf{k}}} E^{\rm pol}
    & = \frac{2\sigma}{N_p} A_{n\mathbf{k}}
    \left( \varepsilon_{n\mathbf{k}} - \varepsilon^{\rm loc} \right) \nonumber \\
    & \quad - \frac{2}{N_p^2}
    \sum_{m\nu\mathbf{q}}
    \big[ A_{m\mathbf{k-q}}
    B^*_{\mathbf{q}\nu} g_{nm\nu}(\mathbf{k-q}, \mathbf{q}) \nonumber \\
    & \qquad +
    A_{m\mathbf{k+q}}
    B_{\mathbf{q}\nu} g^*_{mn\nu}(\mathbf{k}, \mathbf{q})\big], \label{eq:grad-el} \\
    \nabla_{B_{\mathbf{q}\nu}} E^\mathrm{pol}
    & = \frac{2}{N_p} B_{\mathbf{q}\nu} \omega_{\mathbf{q}\nu} \nonumber \\
    & \quad -
    \frac{2}{N_p^2}
    \sum_{nm\mathbf{k}}
    A^*_{m\mathbf{k+q}}
    g_{mn\nu}(\mathbf{k},\mathbf{q})
    A_{n\mathbf{k}}. \label{eq:grad-ph}
\end{align}
These gradients enable optimization of the binding energy using efficient gradient-based techniques~\cite{Vasilchenko2025}.
They also facilitate the identification of multiple polaronic solutions on the pPES within a single system.
Moreover, they allow MEPs between different polaronic configurations to be determined.
In what follows, we show how Eqs.~\eqref{eq:epol}-\eqref{eq:grad-ph} can be used for this purpose through the simplified string method~\cite{E2007}.

\subsection{String method for minimum-energy paths}

A formulation of the string method can be found in Refs.~\cite{E2002, E2007}.
Here, we outline the simplified string method and its algorithmic implementation in the context of the variational polaron equations.

The general problem can be stated as follows: given two polaronic configurations, find the MEP connecting them on the pPES defined by $E^{\rm pol}[\boldsymbol{A},\boldsymbol{B}]$.
The basic idea of the string method is to determine the MEP by evolving an energy curve $\gamma = (\boldsymbol{A},\boldsymbol{B})$, connecting the two polarons, under the potential force field $-\nabla E^{\rm pol}$.
The curve is an MEP if the normal component of the force field vanishes,
\begin{align}
    \nabla E_\perp^{\rm pol} [\gamma] = 0,
\end{align}
where
\begin{equation}
    \nabla E_\perp^{\rm pol} [\gamma] =
    \nabla E^{\rm pol} [\gamma] -
    \nabla E_\parallel^{\rm pol}[\gamma].
\end{equation}
Here, the tangential component is given by
\begin{equation}
    \nabla E_\parallel^{\rm pol}[\gamma] =
    \bigl(\nabla E^{\rm pol}[\gamma] \cdot \hat{\tau} [\gamma] \bigr)
    \hat{\tau} [\gamma],
\end{equation}
where $\hat{\tau}[\gamma]$ denotes the unit tangent vector to the curve.
For a curve represented by a discrete set of points, $\gamma = \{ \boldsymbol{A}_i,\boldsymbol{B}_i \}$, the evolution of each point is governed only by the normal component of the force field, $\nabla E_\perp^{\rm pol}[\gamma]$.
By contrast, the tangential component $\nabla E_\parallel^{\rm pol}[\gamma]$ moves a point along the curve without changing its shape, modifying only the curve parametrization.
The key idea of the simplified string method is therefore to evolve the curve using the full force field $\nabla E^{\rm pol}[\gamma]$ while enforcing a fixed parametrization.

In the variational polaron equations framework, the polaron force field has separate electronic and vibrational components, $\nabla_{A_{n\mathbf{k}}} E^{\rm pol}$ and $\nabla_{B_{\mathbf{q}\nu}} E^\mathrm{pol}$.
Moreover, for a fixed deformation field $\boldsymbol{B}$, the optimal electronic configuration $\boldsymbol{A}$ can always be obtained by minimizing $E^{\rm pol}[\boldsymbol{A},\boldsymbol{B}]$ with respect to the electronic coefficients only.
Hence, an energy curve $\gamma$ can be parametrized solely by the phonon coefficients $\boldsymbol{B}$, since they uniquely determine the electronic coefficients $\boldsymbol{A}$, up to a global phase, through this constrained optimization.
Consequently, the evolution of a discretized curve is governed by the phonon gradient in Eq.~\eqref{eq:grad-ph}, together with a fixed parametrization in terms of the deformation field,
\begin{equation}
    \gamma_{\boldsymbol{B}} : \alpha_i \leftrightarrow \boldsymbol{B}_i, \quad
    \alpha_i \in [0,1].
\end{equation}
This simplifies the search for MEPs to the constrained evolution of the deformation field.

The process can then be expressed as follows.
For a curve represented by an array of polaron images with deformation fields $\boldsymbol{B}_i$ and fixed parametrization $\gamma_{\boldsymbol{B}}$, one first determines the corresponding optimal charge distributions $\boldsymbol{A}_i$.
The deformation field is then evolved over an artificial time step $\Delta t$ according to the full phonon force,
\begin{equation}\label{eq:b-ode}
    \dot{\boldsymbol{B}}_i =
    -\nabla_{\boldsymbol{B}} E^\mathrm{pol}
    [\boldsymbol{A}_i, \boldsymbol{B}_i].
\end{equation}
After this evolution step, the updated deformation fields, $\tilde{\boldsymbol{B}}_i$, may no longer satisfy the chosen parametrization because of the tangential force component.
The parametrization is therefore restored by redistributing the points along the curve, $\tilde{\boldsymbol{B}}_i \to \boldsymbol{B}_i$.
In what follows, we describe the algorithmic details of this procedure.

\subsection{String method: algorithmic details}

The implementation of the string method within the variational polaron equations framework involves the following steps.
Let $\boldsymbol{B}^{(n)}_i$ denote the array of deformation fields corresponding to the $N+1$ polaron images at the $n$-th iteration of the energy-curve evolution.
To update the curve, we first optimize the electronic degrees of freedom $\boldsymbol{A}^{(n)}_i$ for each image at fixed deformation fields $\boldsymbol{B}^{(n)}_i$.
This corresponds to the optimization of a single polaron state at fixed geometry.
The procedure follows the general gradient-based workflow of the variational polaron equations, described in Ref.~\citenum{Vasilchenko2025}.

Next, the integration of Eq.~\eqref{eq:b-ode} can be performed using any suitable solver.
In our implementation, we use the forward Euler method:
\begin{equation} \label{eq:rk-4th}
\tilde{\boldsymbol{B}}^{(n)}_i
=
\boldsymbol{B}^{(n)}_i
-
\Delta t \,
\nabla_{\boldsymbol{B}} E^{\rm pol}[\boldsymbol{A}^{(n)}_i,\boldsymbol{B}^{(n)}_i],
\end{equation}
where the time step $\Delta t$ is taken to be the same for all images.
The updated set of configurations $\boldsymbol{B}^{(n+1)}_i$ is obtained through reparametrization,
\begin{equation}
     \tilde{\boldsymbol{B}}^{(n)}_i \to {\boldsymbol{B}}^{(n+1)}_i.
\end{equation}
In the present implementation, the parametrization is enforced using equal arc length, so that the images remain equidistant along the curve.
To achieve this, the intermediate values $\tilde{\boldsymbol{B}}^{(n)}_i$ are first associated with a non-uniform mesh $\tilde{\alpha}^{(n)}_i$:
\begin{equation}
\begin{aligned}
    \tilde{\alpha}^{(n)}_i &= s_i/s_{N}, \\
    s_i &= s_{i-1} + \|\tilde{\boldsymbol{B}}^{(n)}_i - \tilde{\boldsymbol{B}}^{(n)}_{i-1} \|, \quad 1\le i\le N, \\
    s_0 &= 0.
\end{aligned}
\end{equation}
Interpolation is then performed,
\begin{equation}
    (\tilde{\alpha}^{(n)}_i, \tilde{\boldsymbol{B}}^{(n)}_i) \to
    ({\alpha}_i, \boldsymbol{B}^{(n+1)}_i),
\end{equation}
to obtain the images on the uniform mesh $\alpha_i = i/N$.
For this step, we use cubic spline interpolation.
The process is iterated until the maximum displacement change of the curve,
\begin{equation}\label{eq:string-displ}
    d = \Delta t^{-1}\,\max_i \| \boldsymbol{B}^{(n+1)}_i - \boldsymbol{B}^{(n)}_i \|,
\end{equation}
falls below a chosen tolerance, $d < \delta$.
In this work, we use $\delta = 10^{-6}$.

\subsection{String method: initial configurations}

The final component required for the MEP optimization procedure is the specification of an initial energy path.
For $N+1$ polaron images, this path can be constructed from the deformation fields representing the initial and final configurations, $\boldsymbol{B}_0$ and $\boldsymbol{B}_N$, respectively.

In the simplest scenario, once a polaron with configuration $\boldsymbol{B}_0$ has been obtained by solving the variational polaron equations, $N_p$ translated solutions,
\begin{equation}
    \{ e^{i\mathbf{q} \cdot \mathbf{R}_p}\boldsymbol{B}_0 \},
\end{equation}
can be automatically generated by primitive translations $\mathbf{R}_p$~\cite{Vasilchenko2025}.
Hence, $\boldsymbol{B}_N$ may be chosen from among these translated configurations.
Alternatively, one may solve the variational polaron equations for two translationally inequivalent polarons~\cite{Vasilchenko2025} and use the corresponding deformation fields as $\boldsymbol{B}_0$ and $\boldsymbol{B}_N$.

\begin{figure}[t]
    \includegraphics[width=\linewidth]{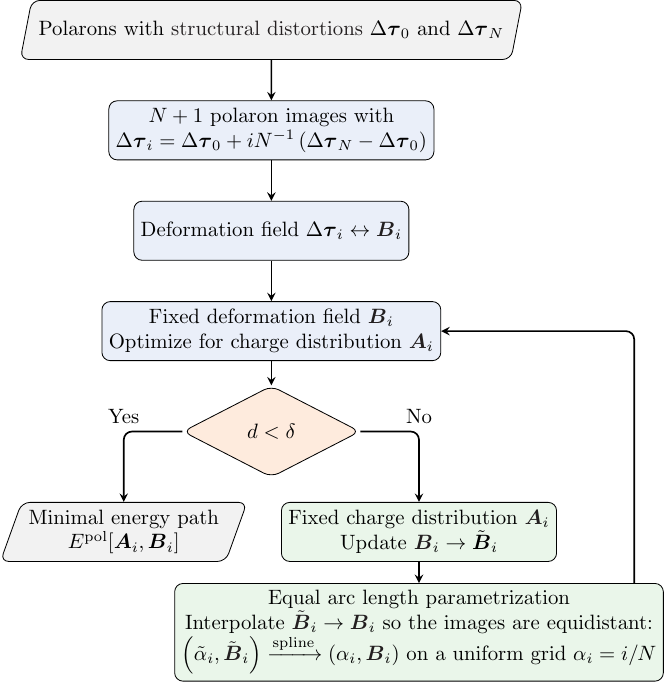}
    \caption{
    {\bfseries Flow diagram for polaron minimum-energy path optimization using the simplified string method within the variational polaron equations framework.}
    $\Delta \tau_0$ and $\Delta \tau_N$ are the initial and final structural distortions.
    $\tilde{\boldsymbol{B}}_i$ indicates intermediate deformation fields without the imposed parametrization.
    }
    \label{fig:flow}
\end{figure}

From the initial and final deformation fields, the corresponding structural distortions can be obtained using Eq.~\eqref{eq:delta-tau}.
The initial distribution of images along the energy path can then be generated by linear interpolation,
\begin{equation}
\Delta \boldsymbol{\tau}_i = \Delta \boldsymbol{\tau}_0 + i N^{-1}
\left(\Delta \boldsymbol{\tau}_N - \Delta \boldsymbol{\tau}_0 \right),
\end{equation}
followed by a reciprocal-space transform $\Delta \boldsymbol{\tau}_i \to \boldsymbol{B}_i$ obtained by inverting Eq.~\eqref{eq:delta-tau}:
\begin{equation}
    B_{\mathbf{q}\nu}
    =
    -\frac{\sigma}{2}\sum_{\kappa\alpha p}
    \Delta \tau^*_{\kappa \alpha p}
    \sqrt{2M_\kappa \omega_{\mathbf{q}\nu}}
    e_{\kappa \alpha \nu} (\mathbf{q})
    e^{i\mathbf{q}\cdot\mathbf{R}_p}.
\end{equation}
In this way, the equal-arc-length parametrization of the curve is already enforced at the initial stage.
With this initial guess, all ingredients required to integrate the simplified string method into the variational polaron equations framework are available.
The workflow for finding minimum-energy paths is illustrated in Fig.~\ref{fig:flow}.

\subsection{String method: computational complexity}

For a hopping process, the computational cost of the algorithm scales with the number of images on the MEP, $N+1$, and the size of the $\mathbf{k/q}$-wavevector mesh needed to represent the image with the most extended polaron configuration.
The cost of one iteration scales as $\mathcal{O}(N_{\mathbf{k}} N_\mathbf{q} N_{n}^2 N_{\nu} N)$, where $N_{\mathbf{k}}$ and $N_{\mathbf{q}}$ denote the wavevector mesh sizes, $N_{n}$ is the number of bands in the polaron wavefunction expansion of Eq.~\eqref{eq:phi-p}, and $N_{\nu}$ is the number of phonon modes.
The limiting factor is usually the mesh size, which defines the BvK supercell hosting the polaron and may be large for images representing extended polarons, e.g., for $[110]$ and $[111]$ hopping in TiO$_2$, see Section~\ref{sec:tio2-polaron}.
For uniform meshes with $N_{\mathbf{k}} = N_{\mathbf{q}} = N_p$, the cost therefore grows quadratically with the number of primitive cells $N_p$ in the BvK supercell.
The prefactor depends on the primitive cell and the polaron wavefunction ansatz, since $N_n$ and $N_{\nu}$ are independent of $N_p$.

This scaling can be compared with that of explicit DFT supercell calculations combined with transition path optimization methods~\cite{Sheppard2008}.
In our approach, the DFT and DFPT inputs are computed only once, using primitive cells and coarse $\mathbf{k/q}$-wavevector meshes.
Then, the dense $\mathbf{q}$-wavevector meshes are reached by interpolation of the perturbed potentials~\cite{Brunin2020a, Brunin2020b} and the dense $\mathbf{k}$-wavevector meshes by non-SCF calculations, see Section~\ref{sec:comp-details}.
For instance, in rutile TiO$_2$, the DFT and DFPT input is computed with a 6-atom primitive cell on an $8\times8\times8$ $\mathbf{q}$-wavevector mesh, while the largest polaron calculation uses the non-SCF and interpolated inputs on a dense $11\times11\times22$ mesh. 
This mesh corresponds to a BvK supercell comprising $N_p = 2662$ primitive cells and $N_{\rm at} = 15972$ atoms.
Explicit calculations for such a supercell would be prohibitively expensive within conventional plane-wave DFT, whose computational cost scales as $\mathcal{O}(N_{\rm at}^3)$, with $N_{\rm at} \propto N_p$~\cite{Goedecker1999}.
Moreover, supercell transition-path calculations require a self-consistent DFT calculation for each of the $N+1$ images at every optimization step.
Localizing the polaron typically requires hybrid functionals, DFT+$U$, or other self-interaction corrections~\cite{Falletta2025}, adding to the computational cost.

The present approach thus reduces the scaling of the cost per path-optimization iteration with BvK supercell size from cubic, $\mathcal{O}(N_p^3)$, to quadratic, $\mathcal{O}(N_p^2)$, for a fixed number of images.
It also allows the primitive-cell DFT and DFPT inputs to be reused across hopping pathways and mesh sizes, with denser meshes accessed through interpolation and non-SCF calculations.

\subsection{Polaron hopping transport}

Once the MEPs are obtained, they can be analysed within Marcus theory for electron transfer in chemical reactions~\cite{Marcus1993}.
In the context of polaron motion, similar theoretical descriptions were developed by Emin and Holstein~\cite{Emin1969}, and by Austin and Mott~\cite{Austin2001}.
For polaron hopping, these approaches are formally equivalent~\cite{Deskins2007} and we refer to them as the Marcus-Emin-Holstein-Austin-Mott (MEHAM) theory.
Within our framework, this theory is applied to describe adiabatic polaron hopping.

\begin{figure}[t]
    \centering
    \includegraphics[width=0.7\linewidth]{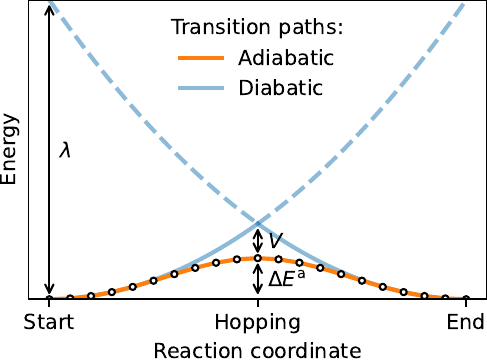}
    \caption{
    {\bfseries Configurational energy diagram for polaron transfer in a representative system.}
    Open circles represent the MEP for polaron hopping obtained using the simplified string method.
    Solid orange and blue curves represent the adiabatic and diabatic transition paths, respectively.
    The latter is obtained by parabolic fitting of the MEP near the initial and final configurations, and the continuations of the fitted parabolas are shown as dashed blue curves.
    The activation energy, electronic coupling, and reorganization energy are denoted by $\Delta E^{\rm a}$, $V$, and $\lambda$, respectively.
    }
    \label{fig:mep}
\end{figure}

The key ingredient in the MEHAM theory is the configurational diagram for the transition between hopping sites.
In our approach, this diagram is obtained from MEP optimization and is schematically illustrated in Fig.~\ref{fig:mep}.
The initial and final hopping states are treated as localized polarons oscillating around their equilibrium lattice configurations.
If a polaron has sufficient energy to overcome the energy barrier, it may transfer from the initial to the final configuration.
Within this theory, the transition may occur in either the adiabatic or diabatic regime.
In the adiabatic case, the electronic coupling between the initial and final states is strong, so that the frequency of electronic transfer is high compared to the characteristic frequency of nuclear motion.
The charge then follows the nuclear motion instantaneously, and the system remains on the lowest adiabatic potential-energy surface.
The dominant transfer mechanism is therefore thermally activated hopping over the activation barrier.
In the diabatic case, the frequency of electronic transfer is lower than the characteristic frequency of atomic vibrations.
The transition then occurs between diabatic potential-energy surfaces and involves electron tunnelling between the initial and final states near their crossing point, with a transfer probability per crossing smaller than unity.

Figure~\ref{fig:mep} shows the adiabatic and diabatic transition paths.
By construction, however, the variational polaron equations formalism assumes the adiabatic strong-coupling approximation.
Therefore, only adiabatic transitions are described consistently within the present approach.
In what follows, we therefore focus on adiabatic polaron transport.

In the MEHAM framework, charge transfer is characterized by three main parameters: the activation energy $\Delta E^{\rm a}$, the electronic coupling $V$, and the reorganization energy $\lambda$.
The activation energy $\Delta E^{\rm a}$ represents the energy barrier for adiabatic transfer.
The electronic coupling $V$ quantifies the interaction between the initial and final polaron wavefunctions and reflects the lowering of the diabatic barrier.
The reorganization energy $\lambda$ is the vertical excitation energy required to transfer the polaron from the initial to the final electronic configuration at fixed lattice deformation.

Typically, the diabatic parameters $\lambda$ and $V$ are estimated by parabolic fitting of the MEP~\cite{Deskins2007}.
Alternatively, the polaron configurations along the MEP allow these parameters to be computed directly, without relying on a fitting model.
This may be important when the MEP is not well described by the standard parabolic model~\cite{Houmsi2025}.
For instance, one can define a transition path corresponding to a vertical excitation of the polaron wavefunction from the initial to the final state, $\phi^{\rm p}_0 \to \phi^{\rm p}_N$, followed by relaxation of the polaron geometry along the reaction coordinate:
\begin{equation}
    \lambda^{0\to N}_i
    =
    E^{\rm pol} [\boldsymbol{A}_N, \boldsymbol{B}_i]
    -
    E^{\rm pol} [\boldsymbol{A}_N, \boldsymbol{B}_N].
\end{equation}
The reorganization energy can then be estimated as $\lambda \equiv \lambda^{0\to N}_0$.

The electronic coupling can be evaluated at the image $i_{\rm hop}$ corresponding to the hopping state as~\cite{Wu2018}:
\begin{equation}
\begin{aligned}  \label{eq:el-coupling}
    V & = H^{0N}_{i_{\rm hop}}, \\
    H^{jk}_{i} & = \langle \phi_j^{\rm p} | \hat{H}^{\rm pol} [\boldsymbol{B}_{i}] | \phi_k^{\rm p} \rangle.
\end{aligned}
\end{equation}
Here, the polaron Hamiltonian is
\begin{multline}
    H^{\rm pol}_{n\mathbf{k}, n'\mathbf{k}'} [\boldsymbol{B}]
     =
     \sigma
     \delta_{n\mathbf{k}, n'\mathbf{k}'} \varepsilon_{n\mathbf{k}}
     \\
     -
     \frac{1}{N_p}
     \sum_{\nu}
     \Bigl[
     B_{\mathbf{k}'-\mathbf{k},\nu}
     g^*_{n'n\nu} (\mathbf{k}, \mathbf{k}'-\mathbf{k}) \\
     +
     B^*_{\mathbf{k}-\mathbf{k}',\nu}
     g_{nn'\nu} (\mathbf{k}', \mathbf{k}-\mathbf{k}')
     \Bigr],
\end{multline}
obtained from Eq.~\eqref{eq:grad-el} by setting $\nabla_{A_{n\mathbf{k}}} E^{\rm pol} \equiv 0$ and imposing the normalization $\langle \phi^{\rm p} | \phi^{\rm p} \rangle = 1$.
This definition of the electronic coupling can be further refined as~\cite{Farazdel1990}
\begin{equation}
    V^{\rm eff} = \label{eq:el-coupling-farazdel}
    \frac{V - s(H^{00}_{i_{\rm hop}} + H^{NN}_{i_{\rm hop}})/2}{1 - s^2},
\end{equation}
where $s = \langle \phi_0^{\rm p} | \phi_N^{\rm p} \rangle$ is the wavefunction overlap.

Adiabatic transfer is also characterized by the attempt frequency $\nu$.
This additional parameter represents the characteristic frequency of nuclear motion associated with the hopping process.
It can be approximated as the frequency of the phonon mode most strongly coupled to polaron formation, such as the longitudinal optical phonon mode at $\mathbf{q}= \Gamma$~\cite{Deskins2007, LafuenteBartolome2024}, $\nu \equiv \nu^{\rm LO}$:
\begin{equation}\label{eq:nu-lo}
    h\nu^{\rm LO} = \omega^{\rm LO}_{\mathbf{q}=\Gamma}.
\end{equation}
Alternatively, within the present variational framework, $\nu$ can be estimated as an effective average of the phonon frequencies at the initial configuration, $\nu \equiv \nu^{\rm eff}$:
\begin{equation}\label{eq:nu-eff}
    h\nu^{\rm eff} = \| \boldsymbol{B}_0 \|^{-2} \sum_{\mathbf{q}\nu} |B_{0,\mathbf{q} \nu}|^2 \omega_{\mathbf{q}\nu}.
\end{equation}
Compared to $\nu^{\rm LO}$, this definition has the advantage of accounting for contributions from multiple phonon modes to adiabatic hopping.
Also, a process-dependent attempt frequency can be estimated from the local curvature of the MEP~\cite{Alkauskas2014},
\begin{equation}\label{eq:e-curvature}
    E(Q_i) - E(Q_{i_0}) \simeq \frac{1}{2}\Omega^2 Q_i^2,
\end{equation}
where
\begin{equation}\label{collective-displ-real}
    Q_i^2 =
    \sum_{\kappa\alpha p} M_{\kappa}
    \left(
        \Delta\tau_{\kappa\alpha p,i}
        - \Delta\tau_{\kappa\alpha p,i_0}
    \right)^2,
\end{equation}
is the collective atomic displacement of image $i$ from the fitting origin, image $i_0$.
Using Eq.~\eqref{eq:delta-tau} and phonon orthonormality gives
\begin{equation}
    Q_i^2 =
    \frac{2}{N_p}\sum_{\mathbf q\nu}
    \frac{
        |B_{i,\mathbf q\nu}-B_{i_0,\mathbf q\nu}|^2
    }{\omega_{\mathbf q\nu}}.
\end{equation}
The corresponding attempt frequency $\nu \equiv \nu^{\rm curv}$ is obtained by fitting Eq.~\eqref{eq:e-curvature} as
\begin{equation}\label{eq:nu-curv}
    \nu^{\rm curv}=\frac{\Omega}{2\pi}.
\end{equation}
In this work, the fit is performed with three images with $i_0 \le i \le i_0 + 2$ or $i_0-2 \le i \le i_0$, if the origin is positioned before and after the transition state, respectively.

For a single-step process, the fitting origin is the initial state, $i_0 = 0$.
For a multi-step process, the MEP must be partitioned into the elementary segments to select $i_0$ for the considered step.
In this work, multi-step processes only occur for the LiF hole polaron, whose hopping MEPs combine on-site rotations with an inter-site hop.
Therefore, we define the hopping segment as the set of images around the hopping transition state above the rotation barrier threshold $\Delta E^{\rm a}_{\rm rot}$.
We then select $i_0$ based on this threshold; see Supplementary Note~S1.
While this definition may better represent an elementary one-dimensional hopping process, it depends on the fitting window and, for multi-step processes, on the chosen threshold.
An alternative route would be to partition the segments and select $i_0$ based on inflection points of the MEP.
However, this might require dense sampling of the MEP with polaronic images and convergence of the local MEP curvature.

A more complete definition of $\nu$ would rely on the Vineyard prefactor~\cite{Vineyard1957}, which takes into account the transverse curvatures of the pPES at the initial state and at the saddle point.
However, such a generalization would require the evaluation of Hessians along the MEPs and lies outside the scope of this work.
We also note that for an elementary one-dimensional hopping process, this generalization reduces to Eq.~\eqref{eq:nu-curv}.

The distinction between the adiabatic and diabatic regimes can be assessed through the transmission coefficient $\kappa$.
Within Landau-Zener theory~\cite{Landau1932a, Landau1932b, Zener1932}, it is given by
\begin{equation}\label{eq:transmission-coefficient}
    \kappa = \frac{2P_{12}}{1 + P_{12}},
\end{equation}
where the Landau-Zener transition probability is
\begin{equation}
    P_{12}
    =
    1
    -
    \exp\left[
    -\frac{\pi^{3/2} |V|^2}{h\nu \sqrt{\lambda k_B T}}
    \right].
\end{equation}
Here $k_B$ is the Boltzmann constant and $T$ is the temperature.
A value of $\kappa$ close to unity indicates adiabatic charge transfer.

In the adiabatic regime, the polaron transfer rate has an Arrhenius-like form,
\begin{equation} \label{eq:k-adiab}
    k^{\rm p, ad} = \nu \exp \Bigl[ -\frac{\Delta E^{\rm a}}{k_{B} T} \Bigr]
\end{equation}
and depends on the activation energy and the effective attempt frequency.
In the diabatic regime, the tunnelling transfer rate is
\begin{equation} \label{eq:k-diab}
     k^{\rm p, diab} =
     \frac{2\pi |V|^2}{\sqrt{4\pi\lambda k_{B} T}}
     \exp \Bigl[ -\frac{(\Delta E^{\rm diff} + \lambda)^2}{4\lambda k_{B} T} \Bigr].
\end{equation}
In this case, the rate depends explicitly on the electronic coupling.
The tunnelling barrier is determined by the reorganization energy and the difference between the binding energies of the final and initial states,
\begin{equation}
    \Delta E^{\rm diff} = E^{\rm pol}[\boldsymbol{A}_N, \boldsymbol{B}_N] - E^{\rm pol}[\boldsymbol{A}_0, \boldsymbol{B}_0].
\end{equation}

The hopping mobility tensor describes the linear response of the polaron drift velocity to a vanishingly small applied electric field, $v_\alpha = \sum_\beta \mu^{\rm p}_{\alpha\beta} \mathcal{E}_\beta$.
In the linear-response regime, the fluctuation-dissipation theorem relates this response to the equilibrium fluctuations of the polaron position in the absence of the field~\cite{Kubo1957, Kubo1966}.
For non-degenerate carriers, such as polarons in the dilute limit, this relation takes the form of the Einstein-Smoluchowski relation~\cite{Einstein1905, Smoluchowski1906},
\begin{equation}\label{eq:mu}
    \boldsymbol \mu^{\rm p} = \frac{e\mathbf{D}}{k_{B} T},
\end{equation}
where the diffusion tensor is evaluated at equilibrium~\cite{Helfand1960},
\begin{equation}\label{eq:diff-tensor}
    D_{\alpha\beta}=\lim_{t\rightarrow\infty} \frac{\langle\Delta r_\alpha(t)\Delta r_\beta(t)\rangle}{2t}.
\end{equation}
Here, $\Delta\mathbf r(t)=\mathbf r(t)-\mathbf r(0)$ denotes the net displacement of the polaron centre during the time interval $t$, and $\langle\dots\rangle$ indicates an ensemble average over trajectories.
In this work, we consider uncorrelated transitions, and Eq.~\eqref{eq:diff-tensor} reduces to~\cite{Dederichs1978}
\begin{equation} \label{eq:diff}
    D_{\alpha\beta} = \frac{1}{2} \sum_j k^\mathrm{p}_{j} R_{j\alpha}R_{j\beta},
\end{equation}
where $j$ runs over the elementary polaron hopping processes from a given site, with transfer rate $k^{\rm p}_{j}$ and displacement vector $\mathbf{R}_j$.
For a single hopping process towards $n_{\rm s}$ equivalent sites at distance $R$, the orientationally averaged diffusion coefficient is obtained from Eq.~\eqref{eq:diff} as
\begin{equation}\label{eq:diff-single}
    \overline{D} = \frac{\operatorname{Tr}\mathbf{D}}{N_{\rm dim}} = \frac{n_{\rm s}R^2k^{\rm p}}{2N_{\rm dim}},
\end{equation}
where $N_{\rm dim}$ is the dimensionality of the material.

Correlations between successive jumps could potentially be included by retaining lattice memory, arising from incomplete relaxation of the lattice distortion after a hop~\cite{Emin1971}.
A possible extension would introduce a time evolution of the deformation field $\boldsymbol{B}(t)$, as in recently developed time-dependent polaron equations~\cite{Garcia-Herrero2026}.
With finite-temperature sampling, such an extension could provide access to correlated hopping trajectories and the diffusion tensor in Eq.~\eqref{eq:diff-tensor}.
Also, it could potentially help to improve the definition of attempt rates beyond the static curvature-based estimates.
This dynamical treatment lies outside the scope of the present work.

Note that MEPs obtained within the present framework correspond to adiabatic transitions because of the underlying adiabatic strong-coupling approximation.
Therefore, the transfer rate is computed with Eq.~\eqref{eq:k-adiab}, assuming $\kappa = 1$, and the corresponding mobility is adiabatic, regardless of the value of the transition probability.
Moreover, the calculations assume independent hopping events and omit dynamical recrossing and correlations between successive jumps.
From Eqs.~\eqref{eq:k-adiab},~\eqref{eq:mu}, and~\eqref{eq:diff-single}, it also follows that the adiabatic mobility of a single hopping process exhibits a crossover from positive to negative temperature dependence at $T^* = \Delta E^{\rm a}/k_B$.
The latter regime corresponds to the saturation of the transfer rate, $k^{\rm p} \to \nu$, and hence of the diffusion coefficient, for $k_B T \gg \Delta E^{\rm a}$, with a tendency towards the $\mu^{\rm p} \sim T^{-1}$ dependence of the mobility.

\section{Data availability}
\label{sec:data}
The data that support the findings of this article are openly available on the Materials Cloud Archive~\cite{matcloud_archive}, DOI:~\href{https://doi.org/10.24435/materialscloud:q2-1c}{10.24435/materialscloud:q2-1c}.

\section{Code availability}
\label{sec:code}
\textsc{ABINIT} is available under the GNU General Public Licence from the \textsc{ABINIT} website: \href{https://www.abinit.org}{https://www.abinit.org}.
The implementation of the string method within the variational polaron equations framework will be made available in the next public release of \textsc{ABINIT}.

\begin{acknowledgments}
V.V. acknowledges funding by the FRS-FNRS Belgium through FRIA.
S. P. is a Research Associate of the Fonds de la Recherche Scientifique - FNRS.
This publication was supported by the Walloon Region in the strategic axis FRFS-WEL-T.
Computational resources have been provided by the supercomputing facilities of the Universit\'e catholique de Louvain (CISM/UCL), the Consortium des Equipements de Calcul Intensif en F\'ed\'eration Wallonie Bruxelles (CECI) funded by the FRS-FNRS under Grant No. 2.5020.11 and computational resources on Lucia, the Tier-1 supercomputer of the Walloon Region with infrastructure funded by the Walloon Region under the Grant Agreement No. 1910247.
\end{acknowledgments}

\section*{Author Contributions}
V.V. formulated and implemented the string method within the variational polaron equations framework, performed the calculations, and wrote the first draft of the manuscript.
M.G. helped with the implementation.
M.G., S.P., and X.G. analysed the results and contributed to writing the manuscript.
S.P. and X.G. supervised the project.
All authors approved the final manuscript.

\section*{Competing Interests}
The authors declare no competing interests.

\newpage
\bibliography{main}

\end{document}


\title{Supplementary Information:\\
Optimal transition states for polaron hopping transport without supercells}

\author{Vasilii Vasilchenko}
\thanks{Corresponding author: vasilii.vasilchenko@uclouvain.be}
\affiliation{European Theoretical Spectroscopy Facility, Institute of Condensed Matter and Nanosciences, Universit\'e catholique de Louvain, Rue de l'Observatoire 8, bte L07.03.01, B-1348 Louvain-la-Neuve, Belgium}
\author{Matteo Giantomassi}
\affiliation{European Theoretical Spectroscopy Facility, Institute of Condensed Matter and Nanosciences, Universit\'e catholique de Louvain, Rue de l'Observatoire 8, bte L07.03.01, B-1348 Louvain-la-Neuve, Belgium}
\author{Samuel Ponc\'e}
\affiliation{European Theoretical Spectroscopy Facility, Institute of Condensed Matter and Nanosciences, Universit\'e catholique de Louvain, Rue de l'Observatoire 8, bte L07.03.01, B-1348 Louvain-la-Neuve, Belgium}
\affiliation{WEL Research Institute, avenue Pasteur 6, 1300 Wavre, Belgium}
\author{Xavier Gonze}
\affiliation{European Theoretical Spectroscopy Facility, Institute of Condensed Matter and Nanosciences, Universit\'e catholique de Louvain, Rue de l'Observatoire 8, bte L07.03.01, B-1348 Louvain-la-Neuve, Belgium}

\date{\today}

\maketitle

\section*{S1. Extraction of curvature-based attempt frequencies}
\label{sec:si-curvature}

The curvature-based attempt frequency $\nu^{\rm curv}$ is obtained by fitting the MEP with a parabola,
\begin{equation}
    E_i - E_{i_0} = \frac{1}{2}\Omega^2 Q_i^2, \qquad \nu^{\rm curv} = \frac{\Omega}{2\pi},
\end{equation}
where $Q_i$ is the mass-weighted displacement of image $i$ from the fitting origin with energy $E_{i_0}$,
\begin{equation}\label{collective-displ-real}
    Q_i^2 =
    \sum_{\kappa\alpha p} M_{\kappa}
    \left(
        \Delta\tau_{\kappa\alpha p,i}
        - \Delta\tau_{\kappa\alpha p,i_0}
    \right)^2.
\end{equation}
In this work, the fit is performed with three images with $i_0 \le i \le i_0 + 2$ or $i_0-2 \le i \le i_0$, if the origin is positioned before and after the transition state, respectively.

\begin{figure}[!htbp]
    \centering
    \includegraphics[width=1.\linewidth]{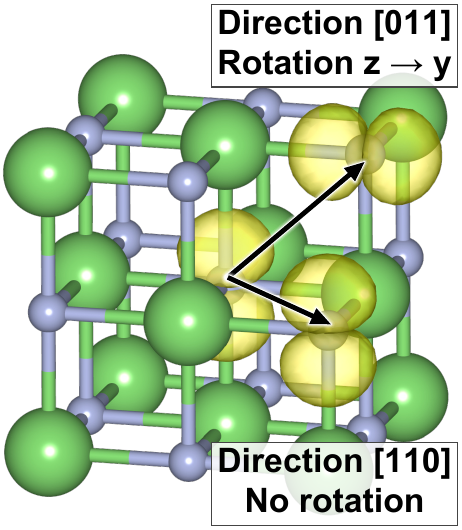}
    \caption{
    {\bfseries Extraction of the curvature-based attempt frequency for hole-polaron hopping in LiF.}
    The hopping MEP corresponds to the $[011]$ transition of a $z$-oriented polaron towards a $z$-oriented polaron, computed with the $12\times12\times12$ $\mathbf{k/q}$-wavevector mesh.
    {\bfseries a}~Hopping MEP including rotation (blue) and pure-rotation MEP (purple), rotation-barrier threshold $\Delta E^{\rm a}_{\rm rot}$ (grey dashed line), and selected hopping origins (green) with their image indices.
    {\bfseries b}~Comparison of the hopping MEP with the pure-rotation MEP as a function of the collective displacement $Q_i$ measured from the left endpoint with $i_0$=0: the pure rotation coincides with the rotation segment that precedes hopping.
    {\bfseries c}~Hopping segment, with $Q_i$ measured from the selected origin, image $i^>_0$=4 of panel {\bfseries a}.
    Red circles identify the images used for the $E(Q_i)-E(Q_{i_0}) = \Omega^2Q_i^2/2$ fit.
    }
    \label{fig:si-curvature}
\end{figure}

For single-step processes, such as the hole-polaron rotation, electron-polaron hopping in LiF and electron-polaron hopping in TiO$_2$, the origin corresponds to the initial MEP endpoint, $i_0 = 0$.
However, the hole-polaron hoppings in LiF are multistep processes, which include a rotation before and after the inter-site hopping step.
The rotation segments of the MEPs must therefore be separated from the pure hopping segments, as illustrated in Fig.~\ref{fig:si-curvature}.

To select the pure hopping segment for extracting $\nu^{\rm curv}_{\rm hop}$, we proceed as follows.
For each $\mathbf{k/q}$-wavevector mesh, we take the activation energy of the pure-rotation MEP, $\Delta E^{\rm a}_{\rm rot}$, as a threshold above which hopping occurs.
On the hopping MEP, we identify the maximum and locate the above-threshold segment, $E > \Delta E^{\rm a}_{\rm rot}$, on both sides of the maximum.
We then choose the adjacent images below the threshold as the fitting origins $i^{>}_0$ and $i^{<}_0$, and compute $Q_i$, $E_i - E_{i^>_0}$ and $E_i - E_{i^<_0}$ from these origins for the next two images towards the transition state.
For instance, for the $12\times12\times12$ $\mathbf{k/q}$-wavevector mesh, $\Delta E^{\rm a}_{\rm rot}=20.39$~meV, and the selection for the hopping of a $z$-oriented polaron to a $z$-oriented polaron along the $[011]$ direction is shown in Fig.~\ref{fig:si-curvature}.

The frequencies $\nu^{\rm curv}_{\rm hop}$ are extracted from both sides of the barrier for all five possible transitions.
For a $z$-oriented polaron, these transitions are $[011]$ hopping towards a final state with $x$, $y$ or $z$ orientation, and $[110]$ hopping towards a final state with $x/y$ or $z$ orientation.
For a fixed $\mathbf{k/q}$-wavevector mesh, the extracted frequencies are averaged, and a linear extrapolation in $N_{p}^{-1/3}$ to the thermodynamic limit is performed.
This yields $\nu^{\rm curv}_{\rm hop} = 5.62$~THz.
%

\clearpage
\section*{S2. Polaron diffusion tensors}
\label{sec:si-diffusion}

In this section, we derive the diffusion tensors of the polarons considered in this work from
\begin{equation}\label{eq:si-diff}
    D_{\alpha\beta} = \frac{1}{2} \sum_j k^\mathrm{p}_{j} R_{j\alpha}R_{j\beta},
\end{equation}
where $j$ runs over the elementary hopping processes from a given site, with transfer rate $k^{\rm p}_{j}$ and displacement vector $\mathbf{R}_j$, as given in the Methods section of the main text.
The mobility tensor then follows from the Einstein-Smoluchowski relation, $\boldsymbol\mu^{\rm p} = e\mathbf{D}/k_BT$.

\textit{Hole polaron in LiF.}
The hole polaron in LiF is localized on fluorine atoms and has three degenerate orientations, each aligned with a Cartesian axis.
Each polaron transfer may involve a rotation of the polaron to achieve a specific orientation before the translation occurs.
We define an elementary pure-hopping event as a transition from an $\alpha$-oriented polaron to a polaron with the same orientation at a nearest-neighbour fluorine site of the face-centred cubic (FCC) lattice with $R_{j\alpha}\ne0$.
For instance, for the $z$-oriented hole, there are eight equivalent transitions with rate $k^{\rm p}_{\rm hop}$ per destination and displacement set
\begin{equation}
    \mathcal R^{(z)}=\frac{R}{\sqrt2} \left\{(\pm1,0,\pm1),(0,\pm1,\pm1)\right\},
\end{equation}
where $R = a/\sqrt{2}$ is the distance to the nearest-neighbour sites, and $a$ is the lattice constant.
This results in diffusion tensors for each orientation that are related by permutation,
\begin{equation}
    \begin{aligned}
        \mathbf D^{(x)} & = R^2k^{\rm p}_{\rm hop}\operatorname{diag}(2,1,1), \\
        \mathbf D^{(y)} & = R^2k^{\rm p}_{\rm hop}\operatorname{diag}(1,2,1), \\
        \mathbf D^{(z)} & = R^2k^{\rm p}_{\rm hop}\operatorname{diag}(1,1,2).
    \end{aligned}
\end{equation}
Since the rotation rate is considerably higher than the hopping rate due to the lower activation energy, $k^{\rm p}_{\rm rot}/k^{\rm p}_{\rm hop} > 10^6$ at 300~K, we assume that on-site rotation equilibrates the three orientations.
Averaging over the orientations yields
\begin{equation}
    \mathbf{D}^{\rm LiF, h} = \frac{4}{3}R^2 k^{\rm p}_{\rm hop} \mathbf{I},
\end{equation}
where $\mathbf{I}$ denotes the identity tensor.

\textit{Electron polaron in LiF.}
The electron polaron in LiF is isotropic and localizes on lithium atoms.
Hopping occurs to the twelve equivalent first-shell nearest-neighbour lithium sites of the FCC lattice with rate $k^{\rm p}_1$ per destination and displacement set
\begin{equation}
    \mathcal R_1 =\frac{R}{\sqrt2}
    \left\{(\pm1,\pm1,0),(\pm1,0,\pm1),(0,\pm1,\pm1)\right\}.
\end{equation}
Thus,
\begin{equation}
    \mathbf{D}^{\rm LiF, e} = 2 R^2 k^{\rm p}_1 \mathbf{I}.
\end{equation}
Only the nearest-neighbour hops are considered here.
However, since the polaron is large, its characteristic length scale $a^{\rm p} = 11.81$~Bohr (6.25~\AA) exceeds the nearest-neighbour hopping distance, $R = 2.87$~\AA.
This suggests that hopping to more distant lithium shells may occur.
Such processes would add hops to the six second-shell and 24 third-shell lithium sites, with displacement sets
\begin{align}
    \mathcal R_2 & =
    \frac{R}{\sqrt2}
    \left\{(\pm2,0,0),(0,\pm2,0),(0,0,\pm2)\right\},
    \\
    \mathcal R_3 & =
    \frac{R}{\sqrt2}
    \left\{(\pm1,\pm1,\pm2),(\pm1,\pm2,\pm1),\right. \nonumber \\
    & \qquad\qquad \left. (\pm2,\pm1,\pm1)\right\}.
\end{align}
If independent elementary transitions to these sites occur with rates $k^{\rm p}_2$ and $k^{\rm p}_3$ per destination, the diffusion tensor becomes
\begin{equation}
    \mathbf{D}^{\rm LiF,e}
    =2R^2\left(k^{\rm p}_1 + k^{\rm p}_2 + 6k^{\rm p}_3\right)\mathbf{I}.
\end{equation}
In this work, these transitions have not been evaluated, and their contribution remains unquantified.

\textit{Electron polaron in rutile TiO$_2$.}
In tetragonal rutile TiO$_2$, the electron polaron localizes on either of the two Ti basis sites with fractional coordinates $(0,0,0)$ and $(1/2,1/2,1/2)$.
For a local polaron orientation with in-plane hopping along $[110]$, we retain two $[001]$ channels, two $[110]$ channels, and eight $[111]$ channels, with displacement sets
\begin{equation}
\begin{aligned}
    \mathcal R_{001} & = \left\{(0,0,\pm c)\right\}, \\
    \mathcal R_{110} &= \left\{(a,a,0),(-a,-a,0)\right\}, \\
    \mathcal R_{111} &= \left\{(\pm a/2,\pm a/2,\pm c/2)\right\},
\end{aligned}
\end{equation}
and rates $k^{\rm p}_{001}$, $k^{\rm p}_{110}$, and $k^{\rm p}_{111}$ per destination.
The $[111]$ hops transfer the polaron between the two Ti basis sites and change its local orientation.
For the symmetry-related local orientation, the in-plane pair is $\{(a,-a,0),(-a,a,0)\}$.
Averaging equally over the two local orientations cancels the off-diagonal $xy$ contributions and gives the anisotropic diffusion tensor
\begin{equation}
    \mathbf{D}^{\rm TiO_2,e}
    =\operatorname{diag}
    \left(D_{ab}, D_{ab}, D_{c}\right),
\end{equation}
with
\begin{equation}
\begin{aligned}
    D_{ab} &= a^2\left(k^{\rm p}_{110} + k^{\rm p}_{111}\right), \\
    D_{c} &= c^2\left(k^{\rm p}_{001} + k^{\rm p}_{111}\right).
\end{aligned}
\end{equation}
%

\clearpage
\section*{S3. Electron polaron dynamics in LiF}
\label{sec:si-lif-electron}

\begin{figure}[!htbp]
    \centering
    \includegraphics[width=\linewidth]{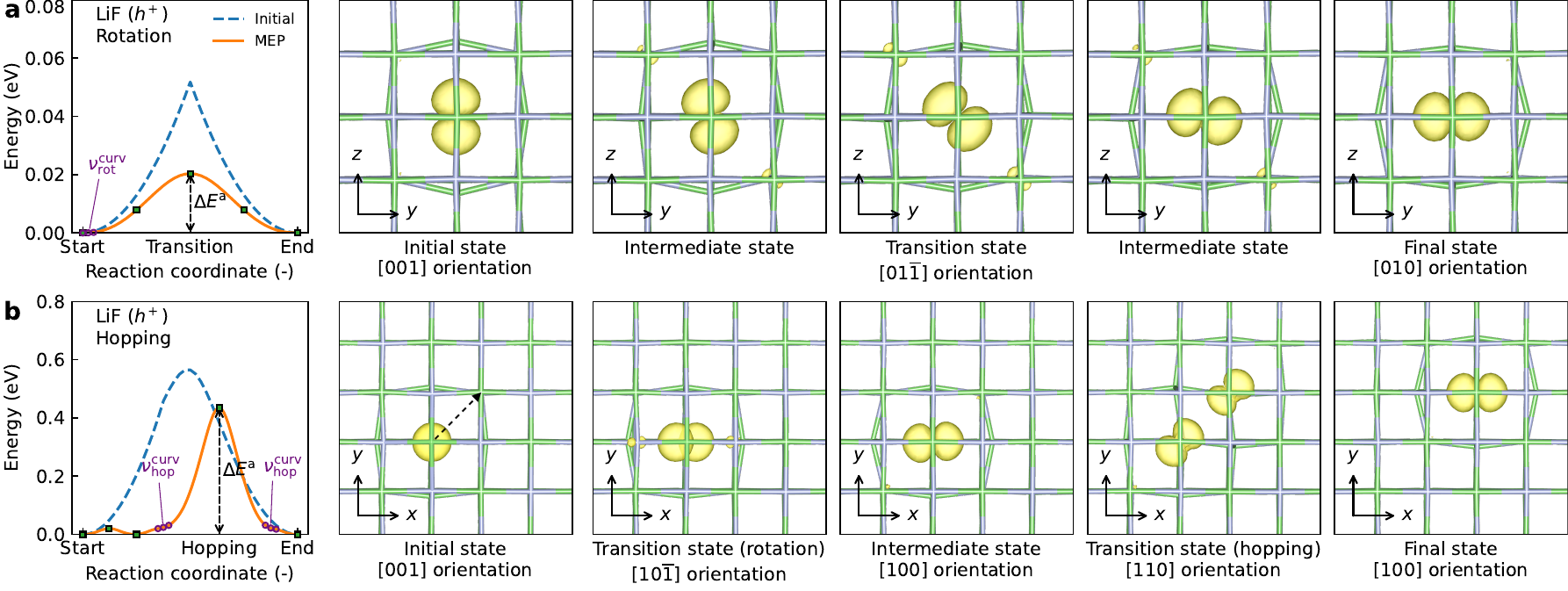}
    \caption{
    {\bfseries Electron polaron dynamics in LiF.}
    The upper left panel shows the initial linearly interpolated energy path and the optimized MEP.
    The other panels show the polaron in the initial, transition, and final configurations, corresponding to the markers on the MEP.
    The activation energy $\Delta E^{\rm a}$ is estimated from the initial state to the transition state.
    The dashed arrow in the initial-state panel indicates the hopping direction.
    }
    \label{fig:si-lif-electron-hopping}
\end{figure}

\begin{figure}[!htbp]
    \centering
    \includegraphics[width=\linewidth]{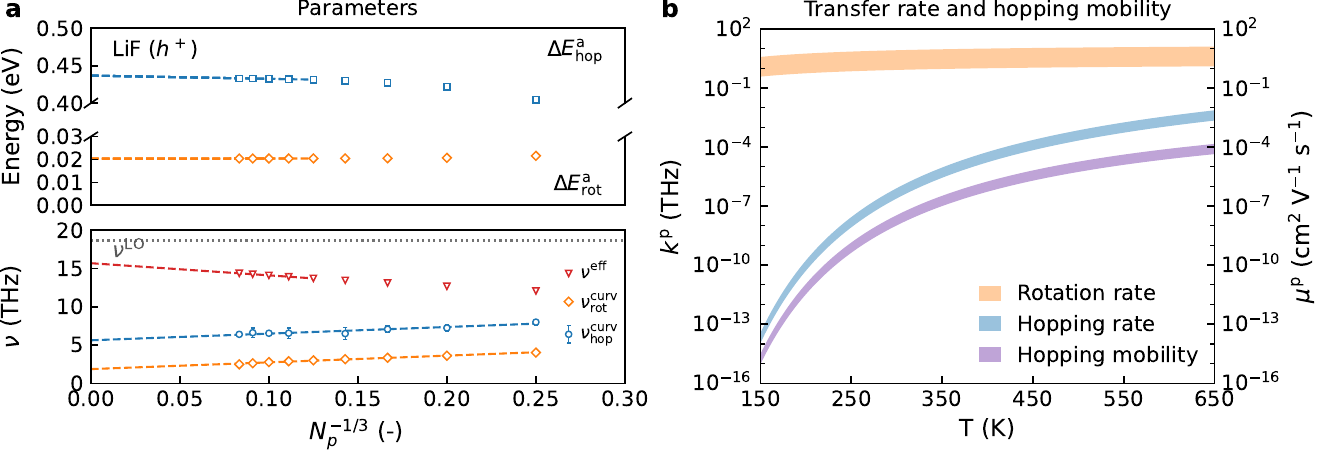}
    \caption{
    {\bfseries Charge-transfer parameters and mobility for the electron polaron in LiF.}
    {\bfseries a}~Activation energy $\Delta E_{\rm hop}^{\rm a}$ for hopping (top) and estimates of the attempt frequency (bottom): phonon-averaged $\nu^{\rm eff}$, curvature-based $\nu^{\rm curv}_{\rm hop}$, and zone-centre LO frequency $\nu^{\rm LO}$.
    The parameters are computed for $\mathbf{k/q}$ wavevector meshes containing $N_p$ wavevectors, of increasing density from $14\times14\times14$ to $20\times20\times20$.
    Dashed lines indicate linear extrapolation to the thermodynamic limit.
    {\bfseries b}~Adiabatic transfer rate $k^{\rm p}$ (left axis) and hopping mobility $\mu^{\rm p}$ (right axis) as a function of temperature, over the range set by the attempt-frequency prefactor: $\nu^{\rm curv}$ and $\nu^{\rm LO}$ define the lower and upper limits.
    }
    \label{fig:si-lif-electron-transport}
\end{figure}

Figure~\ref{fig:si-lif-electron-hopping} shows the optimized MEP for the nearest-neighbour hopping of the electron polaron in LiF, together with the polaron in the initial, transition, and final configurations.
The path optimization lowers the energy barrier by about one order of magnitude with respect to the linearly interpolated path.
The polaron density extends over several lithium shells in all three configurations, which illustrates that the method can be applied numerically to polarons with extended distortions.
The corresponding transition-state theory parameters and transport quantities are shown in Fig.~\ref{fig:si-lif-electron-transport}.
Because of the large polaron size, the calculations require denser $\mathbf{k/q}$-wavevector meshes than for the hole polaron, from $14\times14\times14$ to $20\times20\times20$.
In the thermodynamic limit, the activation energy is $\Delta E^{\rm a} = 0.173$~meV, and the estimates for the attempt frequency are $\nu^{\rm curv}_{\rm hop} =1.50$~THz and $\nu^{\rm eff}=13.08$~THz.
Since $\Delta E^{\rm a} \ll k_B T$ over the whole temperature range, the transfer rate is essentially equal to the attempt frequency, and the mobility decreases with temperature as $T^{-1}$.
As discussed in the main text, these results illustrate the numerical applicability of the framework rather than a validated description of electron transport in LiF.

\clearpage
\section*{S4. Binding and localization energies of the electron polaron in rutile TiO$_2$}
\label{sec:si-tio2-energy}

\begin{figure}[!htbp]
    \centering
    \includegraphics[width=\columnwidth]{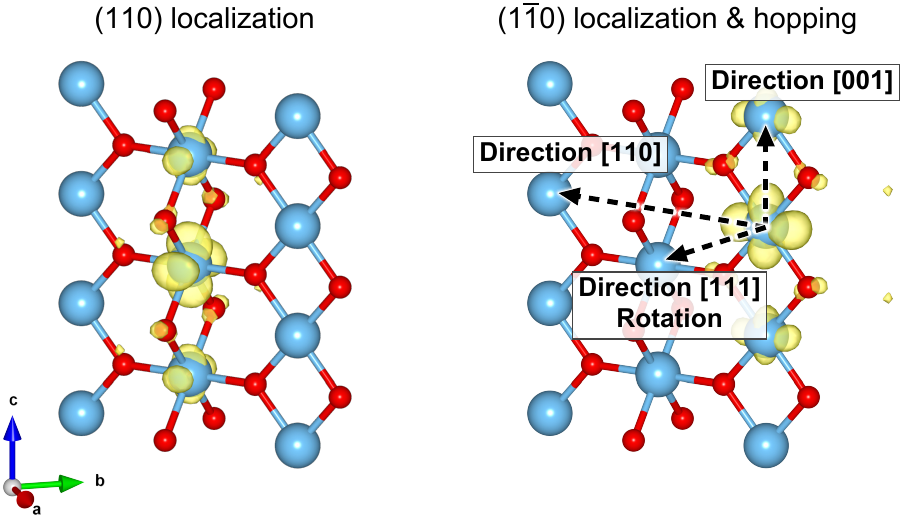}
    \caption{
    {\bfseries Energetics of the electron polaron in rutile TiO$_2$.}
    Convergence of the binding energy $E^{\rm pol}$ and localization energy $\varepsilon^{\rm loc}$ as a function of the inverse linear supercell size $N_p^{-1/3}$.
    The parameters are computed for anisotropic $\mathbf{k/q}$ meshes of increasing density, $n\times n\times 2n$ with $6 \le n \le 11$.
    Dashed lines indicate linear extrapolation to the thermodynamic limit.
    }
    \label{fig:si-tio2-energy}
\end{figure}

Figure~\ref{fig:si-tio2-energy} shows the convergence of the binding energy $E^{\rm pol}$ and localization energy $\varepsilon^{\rm loc}$ of the electron polaron in rutile TiO$_2$ with the size of the BvK supercell.
Both quantities vary linearly with the inverse linear supercell size $N_p^{-1/3}$ for the anisotropic $n\times n\times 2n$ meshes with $6 \le n \le 11$.
The extrapolation to the thermodynamic limit yields $E^{\rm pol}=-0.095$~eV and $\varepsilon^{\rm loc}=-0.714$~eV.
The small magnitude of the binding energy indicates a weakly bound polaron.
%

\newpage

\clearpage
\section*{S5. Comparison of activation barriers in rutile TiO$_2$ with the literature}
\label{sec:si-tio2-barriers}

\begin{figure}[!htbp]
  \centering
  \includegraphics[width=\linewidth]{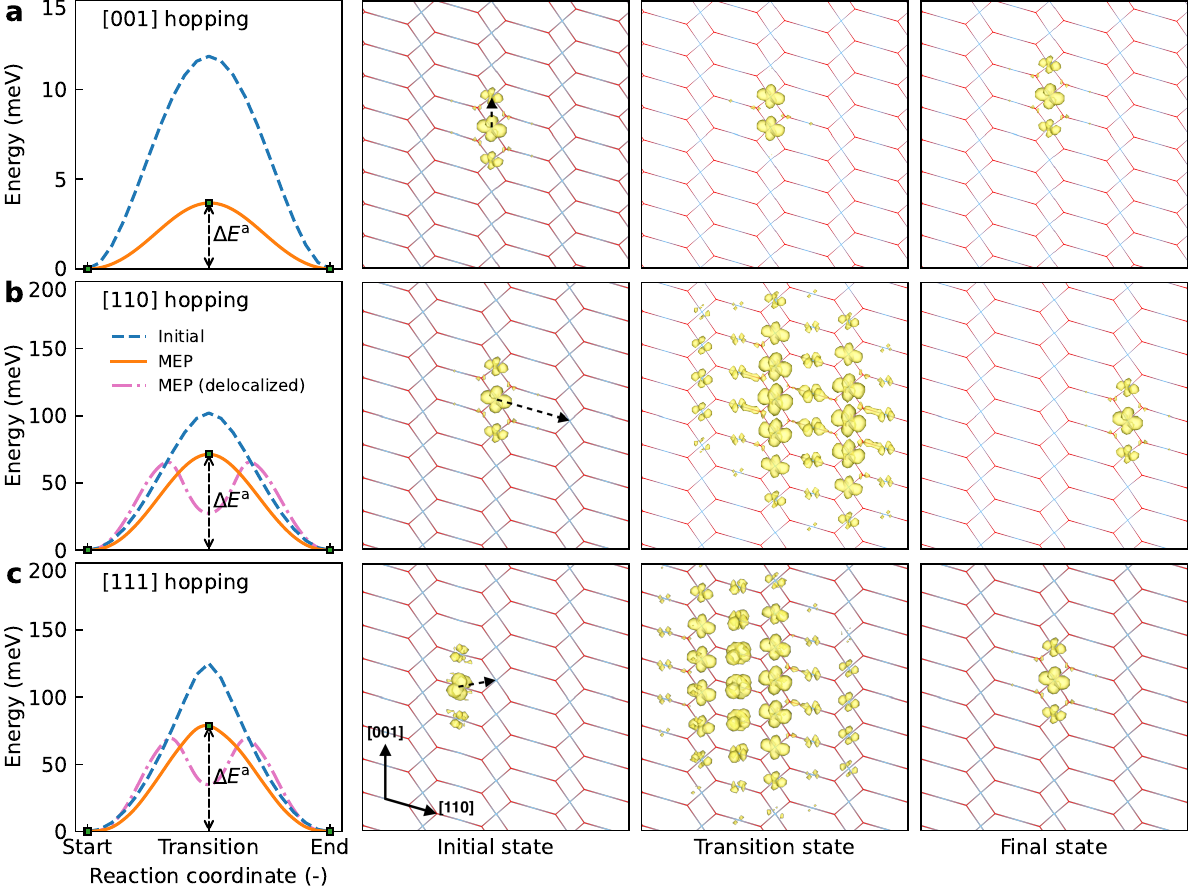}
  \caption{
  {\bfseries Direction-resolved activation energies for electron-polaron hopping in rutile TiO$_2$.}
  The barriers of the optimized MEPs and of the linear paths obtained in this work are compared with results from the \textit{ab initio} polaron equations~\cite{Dai2024c}, DFT+$U$~\cite{Deskins2007, Behara2020}, constrained-DFT valence-bond theory~\cite{Yan2015}, and DeepPolaron MLMD~\cite{Ahart2026}.
  Hatched bars distinguish the three values of $U$ used by Behara and Dupuis~\cite{Behara2020}, and unavailable directional values are omitted.
  The MLMD value is an Arrhenius transport slope, whereas the remaining entries are static or model activation barriers.
  }
  \label{fig:si-tio2-barriers}
\end{figure}

Figure~\ref{fig:si-tio2-barriers} compares the direction-resolved activation energies obtained in this work, for both the optimized MEPs and the initial linearly interpolated paths, with those reported using other theoretical approaches.
Our linear-path barriers along $[001]$ and $[111]$, 12 and 124~meV, agree with those obtained from the \textit{ab initio} polaron equations along linear paths, 13 and 133~meV~\cite{Dai2024c}.
Path optimization reduces them to 3.86 and 78.01~meV, respectively.
DFT+$U$ calculations yield larger barriers, which increase with the Hubbard parameter~\cite{Deskins2007, Behara2020}, while constrained-DFT valence-bond calculations~\cite{Yan2015} give intermediate values.
The MLMD value~\cite{Ahart2026} is an activation energy extracted from the Arrhenius slope of the transport, whereas the other entries are static or model activation barriers, so that they are not strictly equivalent.
All approaches that resolve several directions predict $[001]$ to be the most favourable hopping direction.
%

\newpage

\clearpage
\section*{S6. Band-like mobility of rutile TiO$_2$ from the Boltzmann transport equation}
\label{sec:si-tio2-ibte}

\begin{figure}[!htbp]
    \centering
    \includegraphics[width=\columnwidth]{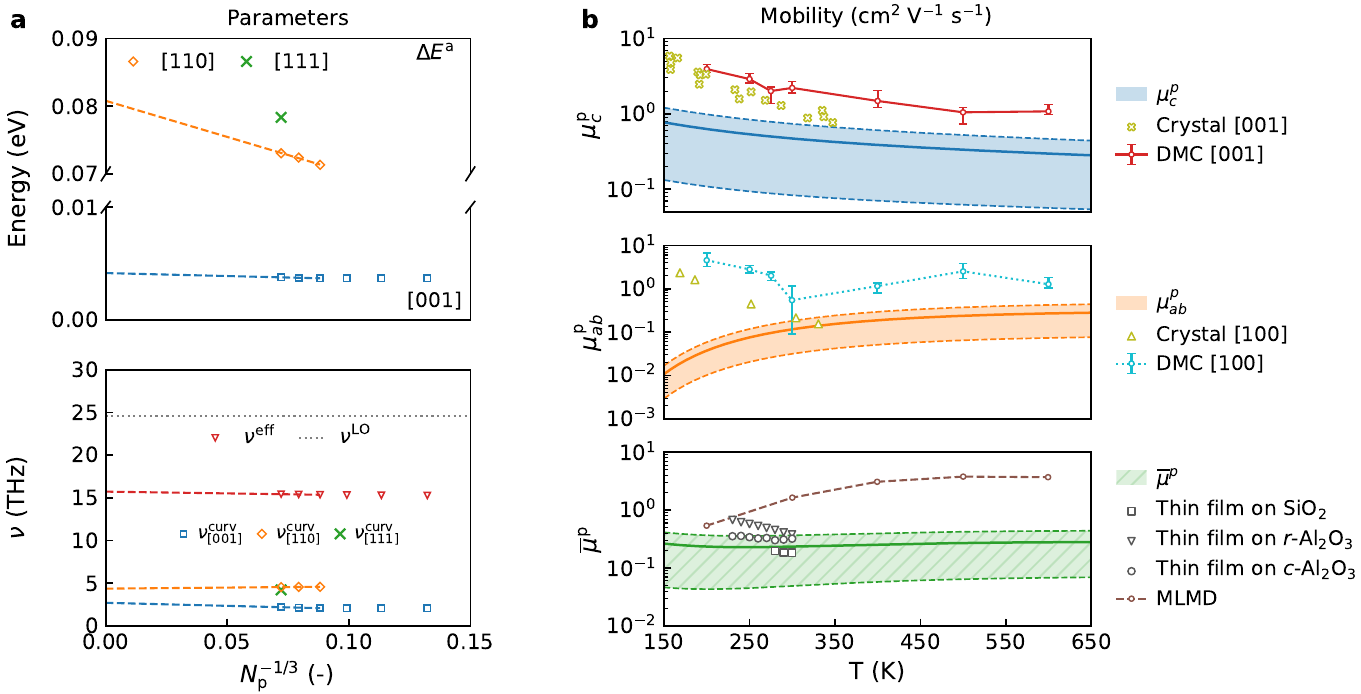}
    \caption{
    {\bfseries Hopping and band-like mobilities of the electron in rutile TiO$_2$.}
    Components {\bfseries a}~$\mu^{\rm p}_c$ and {\bfseries b}~$\mu^{\rm p}_{ab}$ of the hopping mobility tensor, compared with iBTE calculations and with direction-resolved measurements in oxygen-deficient single crystals~\cite{Yagi1996}.
    Shading spans the mobility values obtained using $\nu^{\rm curv}$ and $\nu^{\rm LO}$ as the attempt-frequency prefactor, and the solid curves within the shaded regions indicate the results obtained with $\nu^{\rm eff}$.
    }
    \label{fig:si-tio2-mobility}
\end{figure}

Figure~\ref{fig:si-tio2-mobility} compares the hopping mobility tensor components $\mu^{\rm p}_c$ and $\mu^{\rm p}_{ab}$ with the phonon-limited band-like mobility obtained from the iterative Boltzmann transport equation (iBTE)~\cite{Ponce2020}, as implemented in \textsc{ABINIT}~\cite{Brunin2020a, Brunin2020b, Claes2022}, and with direction-resolved measurements in oxygen-deficient single crystals~\cite{Yagi1996}.
The iBTE mobilities exceed both the hopping mobilities and the measured values along the two directions, the latter by one to two orders of magnitude.
Impurity and defect scattering, which are not included in the iBTE calculations, can account for part of this difference.
However, the size of the overestimation reflects the fact that the iBTE describes band-like transport of delocalized carriers and does not account for the self-trapping of the polaron.
%

\newpage

\clearpage
\onecolumngrid
\section*{S7. Computational parameters}
\label{sec:si-parameters}

\begin{table}[!htbp]
  \centering
  \caption{
    Materials studied in this work, with optimized lattice constants and computational parameters.
    $E^\mathrm{cut}$ denotes the plane-wave energy cutoff.
    The coarse $\mathbf{k}$- and $\mathbf{q}$-wavevector meshes are the minimal $\Gamma$-centred uniform grids required for the convergence of:
    (i) KS electronic bands $\varepsilon_{n\mathbf{k}}$, wavefunctions $\psi_{n\mathbf{k}}$, phonon dispersions $\omega_{\mathbf{q}\nu}$, and eigenmodes $e_{\kappa\alpha\nu} (\mathbf{q})$;
    (ii) dielectric tensor $\boldsymbol{\epsilon}^\infty$ and Born effective charges $\boldsymbol{Z}$;
    (iii) dynamical quadrupoles $\boldsymbol{Q}$.
    For cubic systems with the Fm$\overline{3}$m space group, the quadrupoles vanish by symmetry and are not computed.
    The last column gives the largest dense (interpolated) $\mathbf{k/q}$-wavevector mesh used for the polaron calculations, which defines the BvK supercell.
  }
  \label{tab:systems}
  \setlength{\tabcolsep}{4pt}
  \begin{tabular}{@{}llcrrccccc@{}}
    \hline \hline
    Material & Space group & XC & \multicolumn{2}{c}{Lattice (Bohr)}
      & $E^{\rm cut}$ (Ha) & \multicolumn{3}{c}{Coarse $\mathbf{k/q}$ mesh} & Largest dense $\mathbf{k/q}$ mesh \\
    \cmidrule(lr){4-5}\cmidrule(lr){7-9}
    & & & $a$ & $c$ &
      & (i) & (ii) & (iii) & \\
    \midrule
    \multirow{2}{*}{LiF} & \multirow{2}{*}{$\mathrm{Fm}\overline{3}\mathrm{m}$} & \multirow{2}{*}{PBE} & \multirow{2}{*}{7.6753} & & \multirow{2}{*}{45}
      & \multirow{2}{*}{$6\times6\times6$} & \multirow{2}{*}{$6\times6\times6$} & & $12\times12\times12$ ($h^+$) \\
      & & & & & & & & & $20\times20\times20$ ($e^-$) \\
    TiO$_2$ (rutile) & $\mathrm{P}4_2/\mathrm{mnm}$ & LDA & 8.5878 & 5.5141 & 40
      & $8\times8\times8$ & $8\times8\times8$ & $8\times8\times8$ & $11\times11\times22$ ($e^-$) \\
    \hline \hline
  \end{tabular}
\end{table}

Table~\ref{tab:systems} reports the optimized lattice parameters and the computational parameters of the DFT and DFPT calculations for LiF and rutile TiO$_2$.
The coarse $\mathbf{k}$- and $\mathbf{q}$-wavevector meshes correspond to the ground-state and DFPT calculations performed in the primitive cell.
The electronic quantities on denser $\mathbf{k}$-wavevector meshes are obtained from non-self-consistent calculations, while the phonons and perturbed potentials are interpolated onto denser $\mathbf{q}$-wavevector meshes.
The last column reports the largest dense $\mathbf{k/q}$-wavevector mesh used in the polaron calculations of the main text, which defines the largest BvK supercell hosting the polaron.

\clearpage
\twocolumngrid
\renewcommand{\bibsection}{\section*{References}}
\bibliography{main}